\documentclass{aastex}
\usepackage{spr-astr-addons,natbib}
\usepackage{url}

\RequirePackage{color}

\def\be{\begin{equation}}
\def\ee{\end{equation}}
\def\bea{\begin{eqnarray}}
\def\eea{\end{eqnarray}}

\begin{document}

\title{Exact power series solutions of the structure equations of the
general relativistic isotropic fluid stars with linear barotropic and
polytropic equations of state}
\shorttitle{Exact power series solutions of the structure equations of the
general relativistic stars}
\shortauthors{T. Harko \& M. K. Mak}

\author{T. Harko\altaffilmark{1,2}} \and
\altaffiltext{1}{Department of Physics, Babes-Bolyai University, Kogalniceanu Street,
Cluj-Napoca 400084, Romania.}
\altaffiltext{2}{Department of Mathematics, University College London, Gower Street, London
WC1E 6BT, United Kingdom, E-mail: t.harko@ucl.ac.uk}
\author{M. K. Mak\altaffilmark{3}}
\altaffiltext{3}{Departamento de F\'{\i}sica, Facultad de Ciencias Naturales, Universidad de
Atacama, Copayapu 485, Copiap\'o, Chile, E-mail: mankwongmak@gmail.com}

\begin{abstract}
Obtaining exact solutions of the spherically symmetric general relativistic
gravitational field equations describing the interior structure of an
isotropic fluid sphere is a long standing problem in theoretical and
mathematical physics. The usual approach to this problem consists mainly in
the numerical investigation of the Tolman-Oppenheimer-Volkoff and of the
mass continuity equations, which describes the hydrostatic stability of the
dense stars. In the present paper we introduce an alternative approach for
the study of the relativistic fluid sphere, based on the relativistic mass
equation, obtained by eliminating the energy density in the
Tolman-Oppenheimer-Volkoff equation. Despite its apparent complexity, the
relativistic mass equation can be solved exactly by using a power series
representation for the mass, and the Cauchy convolution for infinite power
series. We obtain exact series solutions for general relativistic dense
astrophysical objects described by the linear barotropic and the polytropic
equations of state, respectively. For the polytropic case we obtain the
exact power series solution corresponding to arbitrary values of the
polytropic index $n$. The explicit form of the solution is presented for the
polytropic index $n=1$, and for the indexes $n=1/2$ and $n=1/5$,
respectively. The case of $n=3$ is also considered. In each case the exact
power series solution is compared with the exact numerical solutions, which
are reproduced by the power series solutions truncated to seven terms only.
The power series representations of the geometric and physical properties of
the linear barotropic and polytropic stars are also obtained.
\end{abstract}



\section{Introduction}

Karl Schwarzschild was the first scientist to find the exact solution of the
Einstein's gravitational field equations describing the interior of a
constant density compact astrophysical object in 1916 \citep{Sch1}. The
search for exact solutions describing static neutral, charged, isotropic or
anisotropic stellar type configurations has continuously attracted the
interests of physicists and mathematicians. A wide range of analytical
solutions of the gravitational field equations describing the interior
structure of the static fluid spheres were found in the past 100 years (for
reviews of the interior solutions of Einstein's gravitational field
equations see \citep{0,1b,2b}). Unfortunately, among these many found
solutions, there are very few exact interior solutions of the field
equations satisfying the required general physical conditions. The criteria
for physical acceptability of an interior solution can be formulated as
follows \citep{1b}: 1) the solutions must be integrated from the regular
origin of the stars. 2) the pressure and the energy density be positive
definite at the origin of the stars. 3) the pressure vanishes at the surface
of the stars. 4) the pressure and the energy density be monotonically
decreasing to the surface of the stars for all radius. 5) causality
requirement is that the speed of sound cannot be faster than the speed of
light inside the stars. 6) the interior metric should be joined continuously
with the exterior Schwarzschild metric. Note that in the field of static
spherically symmetric fluid spheres, an
important bound on the mass-radius ratio for stable general relativistic
stars was obtained in \cite{Bu}, given by $2GM/c^{2}R\leq 8/9$, where $M$ is the mass of the star as
measured by its external gravitational field, and $R$ is the boundary radius
of the star. The Buchdahl bound was generalized to include the presence of the cosmological constant as well as higher dimensions and electromagnetic fields in \citep{MCC,Pi1,Pi2,Pi3}.  

In recent years, many exact solutions of the field equations describing the
interior structure of the fluid stars have been found by assuming the
existence of the anisotropic pressure \citep{B1,B2,B3,B4,A1,A3,A4,A2}. Since
there are three independent field equations representing the stellar model,
after adding the anisotropy parameter to the model, one has more
mathematical freedom, and hence it is easier to solve the field
equations analytically. However, it may be unphysical to assume the
existence of anisotropic stresses. For instance, in a compact star, although
the radial pressure vanishes at the surface of the star, one still could
postulate the tangential pressure to exist. While the latter does not alter
the spherical symmetry, it may create some streaming fluid motions \cite{3}.
Thus, in order to obtain a realistic description of stellar interiors in the
following we assume that the matter content of dense general relativistic
can be described thermodynamically by the energy density $\rho \left(
r\right) $ and the isotropic pressure $p\left( r\right) $. Therefore, from a
mathematical point of view the isotropic stellar models are governed by the
three field equations for four unknowns: the $tt$ and $rr$ components of the
metric tensor $\exp \left[ \nu \left( r\right) \right] $ and $\exp \left[
\lambda \left( r\right) \right] $, the energy density $\rho \left( r\right) $%
, and the pressure $p\left( r\right) $ respectively. Thus, the general
relativistic stellar problem is an underdetermined one. In order to
close the system of field equations an equation of state must be imposed.
Very recently, the isotropic pressure equation was reformulated as a Riccati
equation. By using the general integrability condition for the Riccati
equation proposed in \cite{M1,M2}, an exact non-singular solution of the
interior field equations for a fluid star expressed in the form of infinite
series was obtained in \cite{00}. The astrophysical analysis indicates that
this power series solution can be used as a realistic model for static
general relativistic high density objects, for example neutron stars.

In 1939, Tolman rewrote the isotropic pressure equation as the exact
differential form involving the metric tensor components, subsequently
leading him to obtain the eight analytical solutions of the field equations
\citep{Tolman}. However, in order to ensure not to violate the causality
condition, in the present paper, we do not follow Tolman's approach.
Alternatively, we need one more constraint to close the system of the
equations and to satisfy the causality requirement. Hence in the present paper we assume first that the
matter energy density $\rho \left( r\right) $ and the thermodynamic pressure $p\left( r\right) $ obey the
linear barotropic equation of state given by%
\begin{equation}
p\left( r\right) =\gamma \rho \left( r\right) c^{2},  \label{0}
\end{equation}%
where $\gamma $ is the arbitrary constant satisfying the inequality $0\leq
\gamma \leq 1$.  A static interior solution of the field equations in isotropic coordinates with the
equation of state (\ref{0}) was presented in \cite{MMH}. The structure and
the stability of relativistic stars with the equation of state (\ref{0}) were
studied in \cite{PH}. An exact analytical solution describing
the interior of a charged strange quark star satisfying the MIT bag model
equation of state $3p=\rho c^{2}-4B$, where $B$ is a constant, was found in
\cite{BV} under the assumption of spherical symmetry and the existence of a
one-parameter group of conformal motions.

Numerical solutions of Einstein's field equation describing static,
spherically symmetric conglomerations of a photon gas, forming so-called
photon stars, were obtained in \cite{Sch}. The solutions imply a back
reaction of the metric on the energy density of the photon gas. In \cite{Glen} it was pointed out that a class of objects called
Radiation Pressure Supported Stars (RPSS) may exist even in Newtonian
gravity. Such objects can also exist in standard general relativity, and
they are called "Relativistic Radiation Pressure Supported Stars" (RRPSS). The
formation of RRPSSs can take place during the continued gravitational
collapse. Irrespective of the details of the contraction process, the
trapped radiation flux should attain the corresponding Eddington value at
sufficiently large $z>>1$. On the basis of Einstein's theory of relativity,
the principle of causality, and Le Chatelier's principle, in \cite{Ruf} it
was established that the maximum mass of the equilibrium configuration of a
neutron star cannot be larger than $3.2M_{\odot}$. To obtain this result it
was assumed that for high densities the equation of state of matter is given
by $p=\rho c^2$. The absolute maximum mass of a neutron star provides a
decisive method of observationally distinguishing neutron stars from black
holes.

There is a long history in the context of physics and astrophysics for the
study of the polytropic equation of state, defined as \citep{Hor}
\begin{equation}
p\left( r\right) =K\rho ^{\Gamma }\left( r\right) .
\end{equation}%
Here  $K$ is the polytropic constant, and the
adiabatic index $\Gamma $ is defined as $\Gamma =1+1/n$, where $n$ is the
polytropic index. Using the polytropic equation of state, the physicists
have investigated the properties of the astrophysical objects in
Newtonian gravity.  Note that $K$ is fixed in the degenerate system for
instance a white dwarf or a neutron star and free in a non-degenerate
system. The hydrostatic equilibrium structure of a polytropic star is
governed for spherical symmetry by the Lane-Emden equation \citep{Hor}
\begin{equation}
\frac{1}{x^{2}}\frac{d}{dx}\left( x^{2}\frac{dy}{dx}\right) +y^{n}=0,
\label{F1}
\end{equation}%
where the dimensionless variables $y$ and $x$ are defined as
\begin{equation}
x^{2}=\frac{4\pi G\rho _{c}^{\frac{n-1}{n}}}{\left( 1+n\right) K}r^{2},y^{n}=%
\frac{\rho }{\rho _{c}},
\end{equation}%
respectively where $\rho _{c}$ and $r$ are the central density and the
radius of the star, respectively, $G$ is the Newtonian gravitational
constant, and $y$ is the dimensionless gravitational potential. The
Lane-Emden Eq. (\ref{F1}) was first introduced by  \cite{L} and later
studied by  \cite{E}, \cite{Fow} and \cite{Milne}, respectively. In order to ensure the regularity of the solution
at the center of the sphere, Eq. (\ref{F1}) must be solved with the initial
conditions given by
\begin{equation}
y\left( 0\right) =1,\left( \frac{dy}{dx}\right) _{x=0}=0.
\end{equation}%
It is well-known that the exact analytical solutions of Eq. (\ref{F1}) can
only be obtained for $n=0,1,5$ \citep{Hor,1s}. However, not all solutions of
Eq. (\ref{F1}) for $n=5$ were known until the year 2012, when all real
solutions of Eq.~(\ref{F1}) for $n=5$ were obtained in terms of Jacobian and
Weierstrass elliptic functions \citep{PM}.  Two integrable classes of the Emden-Fowler equation of the type $z\rq{}\rq{}=A \chi ^{-\lambda-2}z^n$ for $\lambda=\frac{n-1}{2}$, and $\lambda=n+1$ were discussed in \cite{Mancas}. By using particular solutions of the Emden-Fowler equations both classes were reduced to the form $\ddot \nu +a \dot \nu +b(\nu-\nu^n)=0$, where $a$, $b$ depend only on $\lambda $, and $n$, respectively. For both cases the solutions can be represented in a closed parametric form, with some values of $n$ yielding Weierstrass elliptic solutions.
It is generally accepted that the
power series method is one of the powerful techniques in solving ordinary
differential equations. Thus  the Lane-Emden Eq.~(\ref{F1}%
) was solved by using a power series method in \citep{CA,IW,CH,MN}, where the convergence of the
solutions was also studied.

The polytropic equation of state has also been adopted to study the interior
structure of the fluid stars in the
framework of general relativity \citep{Tooper}. The solution
of the gravitational field equations for relativistic static spherically
symmetric stars in minimal dilatonic gravity using the
polytropic equation of state was presented in \cite{PK}. The general
formalism to model polytropic general relativistic stars with the
anisotropic pressure was considered in \cite{LW}, and its stellar
applications were also discussed. By solving the Tolman-Oppenheimer-Volkoff
(TOV) equation, a class of compact stars made of a charged perfect fluid
with the polytropic equation of state was analyzed in \cite{jj}. Exact
solutions of the Einstein-Maxwell equation with the anisotropic pressure and
the electromagnetic field in the presence of the polytropic equation of
state were obtained in \cite{PS}. Charged polytropic stars, and a
generalization of the Lane-Emden equation was investigated in \cite{RM}.
Using the power series methodology, a new analytical solution of the TOV
equation for polytropic
stars was presented in \cite{MAS}. The divergence and the convergence
of the power series solutions for the different values of the polytropic
index $n$ were also discussed. The gravitational field
equations for the static spherically symmetric perfect fluid models with the
polytropic equation of state can be written as two complementary 3 dimensional
regular systems of ordinary differential equations on compact state space.
Due to the highly nonlinear structure of the systems, it is difficult to
solve them exactly, and thus they were analyzed numerically and qualitatively
using the theory of dynamical systems in \citep{UN,KCC}. The three-dimensional
perfect fluid stars with the polytropic equation of state, matched to the exterior three-dimensional black hole geometry of Ba\~{n}ados,
Teitelboim and Zanelli were considered in \cite{PTA}. A new class of exact solutions for
a generic polytropic index was found, and analyzed. The
structure of the relativistic polytropic stars and the stellar stability
analysis embedded in a chameleon scalar field was discussed in \cite{VD}. In
\cite{XY} a polytropic quark star model was suggested in order to establish
a general framework in which theoretical quark star models could be tested
by the astrophysical observations. Spherically symmetric static matter
configurations with the polytropic equation of state for a class of $f\left(
R\right) $ models in Palatini formalism were investigated in \cite{GO}, and
it was shown that the surface singularities are not physical in the case of
Planck scale modified Lagrangians.

It is the purpose of the present paper to study the interior structure of
the general relativistic fluid stars with the linear barotropic and the
polytropic equations of state, and to obtain exact power series solutions of
the corresponding equations. As a first step in our study we introduce the
basic equation describing the interior mass profile of a relativistic star,
and which we call the relativistic mass equation. This equation is obtained
by eliminating the energy density between the mass continuity equation and
the hydrostatic equilibrium equation. Despite its apparent mathematical
complexity, the relativistic mass equation can be solved exactly for both
linear barotropic and polytropic equations of state, by looking to its exact
solutions as represented in the form of power series. In order to obtain
closed form representations of the coefficients we use the Cauchy
convolution of the power series. In this way we obtain the exact series
solutions for relativistic spheres described by linear barotropic equations
of state with arbitrary $\gamma $, and for the polytropic equation of state
with arbitrary polytropic index $n$. The case $n=1$ is investigated
independently, and the corresponding power series solution is also obtained.
We compare the truncated power series solutions containing seven terms only
with the exact numerical solution of the TOV and mass continuity equations.
In all considered cases we find an excellent agreement between the power
series solution, and the numerical one.

The present paper is organized as follows. The gravitational field
equations, their dimensionless formulation and the basic relativistic mass
equation are presented in Section~\ref{sect1}. The definition of the Cauchy
convolution for infinite power series is also introduced. The non-singular
power series solution for fluid spheres described by a linear barotropic
equation of state is presented in Section~\ref{sect2}. The comparison
between the exact and numerical solutions are presented. The exact power
series solutions for a general relativistic polytropic star with polytropic
index $n=1$ are derived in Section~\ref{sect3}, and the comparison with the
exact numerical solution is also performed. The case of the arbitrary
polytropic index $n$ is considered in Section~\ref{sect4}. The power series
solutions are compared with the exact numerical solutions for the cases $%
n=1/2$, $n=1/5$ and $n=3$, respectively. We discuss our results and conclude
our paper in Section~\ref{sect5}. The first seven coefficients of the power series solution of the
relativistic mass equation for arbitrary polytropic index $n$ are presented in Appendix~\ref{app}.

\section{The gravitational structure equations, dimensionless variables, and
the relativistic mass equation}

\label{sect1}

We start our study by writing down the gravitational field equations
describing a static spherically symmetric general relativistic star, and
presenting the corresponding structure equations for stellar type objects.
In order to simplify the mathematical and the numerical formalism, we
rewrite the basic equations in a set of dimensionless variables, and we
obtain the basic non-linear second order differential equation describing
the mass distribution inside the relativistic stars.

\subsection{Gravitational field equations and structure equations for
compact spherically symmetric objects}

The static and spherically symmetric metric for describing a gravitational
relativistic sphere in Schwarzchild coordinates is given by the line element
\begin{equation}
ds^{2}=e^{\nu }c^{2}dt^{2}-e^{\lambda }dr^{2}-r^{2}d\Omega ^{2},  \label{1n}
\end{equation}%
where the metric components $\nu $ and $\lambda $ are function of radial
coordinate $r$, for simplicity we have denoted the quantity $d\Omega ^{2}$
as $d\Omega ^{2}=d\theta ^{2}+\sin ^{2}\theta d\phi ^{2}$. The Einstein's
gravitational field equations are
\begin{equation}
R_{i}^{k}-\frac{1}{2}R\delta _{i}^{k}=\frac{8\pi G}{c^{4}}T_{i}^{k},
\label{2}
\end{equation}%
where $G$ is the Newtonian gravitational constant, and $c$ is the speed of
light, respectively. For an isotropic spherically symmetric matter
distribution the components of the energy-momentum tensor are of the form
\begin{equation}
T_{i}^{k}=\left( \rho c^{2}+p\right) u_{i}u^{k}-p\delta _{i}^{k},  \label{3}
\end{equation}%
where $u^{i}$ is the four velocity, given by $u^{i}=\delta _{0}^{i}$, and
the quantities $\rho \left( r\right) $ and $p\left( r\right) $ are the
energy density and the isotropic pressure, respectively. For any physically
acceptable stellar models, we require that the energy density and the
pressure must be positive and finite at all points inside the fluid spheres.
By inserting Eqs.~(\ref{1n}) and (\ref{3}) into Eq.~(\ref{2}), the latter
equations yield the Einstein's gravitational field equations describing the
interior of a static fluid sphere as \citep{Landau}
\begin{equation}
-\frac{1}{r^{2}}\frac{d}{dr}\left( re^{-\lambda }\right) +\frac{1}{r^{2}}=%
\frac{8\pi G}{c^{2}}\rho \left( r\right) ,  \label{f1}
\end{equation}%
\begin{equation}
\frac{e^{-\lambda }}{r}\frac{d\nu }{dr}+\frac{e^{-\lambda }-1}{r^{2}}=\frac{%
8\pi G}{c^{4}}p\left( r\right) ,  \label{f2}
\end{equation}%
\bea
&&e^{-\lambda }\Bigg[ \frac{1}{2}\frac{d^{2}\nu }{dr^{2}}+\frac{1}{4}\left(
\frac{d\nu }{dr}\right) ^{2}-\frac{1}{4}\frac{d\nu }{dr}\frac{d\lambda }{dr}+\nonumber\\%
&&\frac{1}{2r}\left( \frac{d\nu }{dr}-\frac{d\lambda }{dr}\right) \Bigg] =%
\frac{8\pi G}{c^{4}}p\left( r\right) .  \label{f3}
\eea
The conservation of the energy-momentum tensor gives the relation%
\begin{equation}
\frac{d\nu }{dr}=-\frac{2}{\rho (r)c^{2}+p(r)}\frac{dp}{dr}.  \label{f4}
\end{equation}%
Eq.~(\ref{f1}) can be immediately integrated to give
\begin{equation}
e^{-\lambda }=1-\frac{2GM(r)}{c^{2}r},  \label{lambda}
\end{equation}%
where $M(r)$ is the mass inside radius $r$. An alternative description of
the interior of the star can be given in terms of the TOV and of the mass
continuity equations, which can be written as
\begin{equation}
\frac{dp}{dr}=-\frac{\left( G/c^{2}\right) \left( \rho c^{2}+p\right) \left[
\left( 4\pi /c^{2}\right) pr^{3}+M\right] }{r^{2}\left( 1-2GM/c^{2}r\right) }%
,  \label{tov}
\end{equation}%
\begin{equation}
\frac{dM}{dr}=4\pi \rho r^{2},  \label{mc}
\end{equation}%
respectively. The system of the structure equations of the star must be
integrated with the initial and boundary conditions
\begin{equation}
M(0)=0,p\left( R\right) =0,
\end{equation}%
where $R$ is the radius of the star, and together with an equation of state $%
p=p\left( \rho \right) $.

\subsection{Dimensionless form of the structure equations}

By introducing a set of dimensionless variables $\eta $ (dimensionless
radial coordinate), $\epsilon \left( \eta \right) $ (energy density), $
P(\eta )$ (pressure) and $m\left( \eta \right) $ (mass), by means of the
transformations
\begin{equation}
r=\eta R,\rho =\rho _{c}\epsilon \left( \eta \right) ,p=\rho
_{c}c^{2}P\left( \eta \right) ,M=M^{\ast }m\left( \eta \right) ,  \label{aa}
\end{equation}%
where $\rho _{c}$ is the central density, the TOV and the mass continuity
equations take the form
\begin{equation}
\frac{dP}{d\eta }=-\frac{a\left[ \epsilon \left( \eta \right) +P\left( \eta
\right) \right] \left[ P\left( \eta \right) \eta ^{3}+m\left( \eta \right) %
\right] }{\eta ^{2}\left[ 1-2am\left( \eta \right) /\eta \right] },
\label{tovad}
\end{equation}%
\begin{equation}
\frac{dm}{d\eta }=\eta ^{2}\epsilon \left( \eta \right) ,  \label{mcad}
\end{equation}%
respectively, where we have fixed the constants $a$ and $M^{*}$ by the
relations
\begin{equation}
a=\frac{4\pi G\rho _{c}}{c^{2}}R^{2},M^{\ast }=4\pi \rho _{c}R^{3}.
\end{equation}

In order to close the above system of equations the dimensionless form $%
P=P(\epsilon )$ of the matter equation of state must also be given. Then the
coupled system of Eqs.~(\ref{tovad}) and (\ref{mcad}) must be solved with
the initial and boundary conditions $\epsilon (0)=1$, $m(0)=0$, and $%
\epsilon \left( 1\right) =\epsilon _{S}$, where $\epsilon _{S}$ is the value
of the surface density of the star, and $\eta =1$ is the value of the
dimensionless radial coordinate $\eta $ on the star's surface. As a function
of the parameter $a$ the radius $R$ and the total mass $M_{S}$ of the star
are given by the relations
\bea
R&=&\sqrt{a}\frac{c}{\sqrt{4\pi G\rho _{c}}}=\nonumber\\
&&\sqrt{a}\times 10.3622\times
\left( \frac{\rho _{c}}{10^{15}\;\mathrm{g/cm^{3}}}\right) ^{-1/2}\;\mathrm{%
km},
\eea
\bea
M_{S}&=&a^{3/2}\frac{c^{3}}{\sqrt{4\pi \rho _{c}G^{3}}}m(1)=\nonumber\\
&&a^{3/2}\times
6.9910\times \left( \frac{\rho _{c}}{10^{15}\;\mathrm{g/cm^{3}}}\right)
^{-1/2}\times m(1)\;M_{\odot },\nonumber\\
\eea
where the quantity $M_{\odot }$ is the mass of the sun. For the mass-radius
ratio of the star, we obtain
\begin{equation}
\frac{GM_{S}}{c^{2}R}=am(1).
\end{equation}

\subsection{The relativistic mass equation and the Cauchy convolution}

By eliminating the energy density $\epsilon (\eta )$ between the equations (%
\ref{mcad}) and (\ref{tovad}) we obtain the following second order
differential equations, which in the following we will call \textit{the
relativistic mass equation},
\bea\label{meq}
&&\hspace{-0.75cm}\eta \frac{d^{2}m(\eta )}{d\eta ^{2}}-2\frac{dm(\eta )}{d\eta }+\nonumber\\
&&\hspace{-0.75cm}\frac{a\left[
\eta ^{3}P\left( \frac{m^{\prime }(\eta )}{\eta ^{2}}\right) +m(\eta )\right]
\left[ m^{\prime }(\eta )+\eta ^{2}P\left( \frac{m^{\prime }(\eta )}{\eta
^{2}}\right) \right] }{\eta \left[ 1-2am(\eta )/\eta \right] P^{\prime
}\left( \frac{m^{\prime }(\eta )}{\eta ^{2}}\right) }=0,  \nonumber\\
\eea
where we have used the simple mathematical relation $dP/d\eta =\left(dP/d\epsilon\right)\left(d\epsilon /d\eta\right)$,
and we have denoted $P^{\prime }\left( \frac{m^{\prime }(\eta )}{\eta ^{2}}\right) =\left.
\frac{dP(\epsilon )}{d\epsilon }\right\vert _{\epsilon =m^{\prime}(\eta )/\eta ^ 2}$.

Equivalently, the relativistic mass equation takes the form
\begin{eqnarray}\label{p4}
\hspace{-0.5cm}&&\left[ 1-\frac{2am(\eta )}{\eta }\right] \left[ \eta \frac{d^{2}m(\eta )}{%
d\eta ^{2}}-2\frac{dm(\eta )}{d\eta }\right] P^{\prime }\left[ \frac{%
m^{\prime }(\eta )}{\eta ^{2}}\right] +\nonumber\\
\hspace{-0.5cm}&&a\eta ^{2}\left\{ \frac{m^{\prime
}(\eta )}{\eta ^{2}}+P\left[ \frac{m^{\prime }(\eta )}{\eta ^{2}}\right]
\right\} \times \Bigg\{ \frac{m(\eta )}{\eta }+\nonumber\\
\hspace{-0.5cm}&&\eta ^{2}P\left[ \frac{m^{\prime }(\eta )}{%
\eta ^{2}}\right] \Bigg\} =0.
\end{eqnarray}

Eq.~(\ref{meq}) must be integrated with the initial conditions $m(0)=0$, and
$m^{\prime }(0)=0$, respectively, and together with the equation of state of
the matter, $P=P(\epsilon )=P\left[ m^{\prime }(\eta )/\eta ^{2}\right] $. It is important to note that the point $\eta =0$ is an ordinary point for Eq.~(\ref{p4}). This is due to the fact that all coefficients in the equation take finite values at the origin. Thus, $\lim _{\eta \rightarrow 0}m(\eta )/\eta=\lim _{\eta \rightarrow 0} \epsilon (\eta)\eta ^3/\eta =0$, and $\lim _{\eta \rightarrow 0}m'(\eta)/\eta ^2=\epsilon (0)=1$, respectively. Since the thermodynamic parameters of the star must be finite at the origin, it follows that $P\left[ \frac{m^{\prime }(\eta )}{\eta ^{2}}\right]$ and $P'\left[ \frac{m^{\prime }(\eta )}{\eta ^{2}}\right]$ are all finite at $\eta =0$.

In the next Sections we will investigate the possibility of obtaining exact
power series solutions of Eq.~(\ref{meq}) for the linear barotropic and the
polytropic equations of state. In order to obtain our solutions we will use
the Cauchy convolution of the power series, defined as follows.

\textbf{Definition}. Let
\bea
f_{1}&=&\sum_{i_{1}=0}^{\infty
}a_{1,i_{i}}x^{i_{1}}, f_{2}=\sum_{i_{2}=0}^{\infty }a_{2,i_{2}}x^{i_{2}}
, \nonumber\\
f_{3}&=&\sum_{i_{3}=0}^{\infty }a_{3,i_{3}}x^{i_{3}}, ...,
f_{s}=\sum_{i_{s}=0}^{\infty }a_{s,i_{s}}x^{i_{s}},
\eea
 be $s$ convergent power
series, $s\geq 2$. Then we define the Cauchy product (convolution) of the $s$
power series, $s\geq 2$, as
\begin{eqnarray}
\hspace{-0.5cm}&&f_{1}\circ f_{2} =\left( \sum_{i_{1}=0}^{\infty
}a_{1,i_{1}}x^{i_{1}}\right) \left( \sum_{i_{2}=0}^{\infty
}a_{2,i_{2}}x^{i_{2}}\right) =\nonumber\\
\hspace{-0.5cm}&&\sum_{i_{1},i_{2}=0}^{\infty }{%
a_{1,i_{1}}a_{2,i_{2}}x^{i_{1}+i_{2}}} = \nonumber\\
\hspace{-0.5cm}&&\sum_{j_{2}=0}^{\infty }\left( \sum_{i_{1}=0}^{j_{2}}{%
a_{1,i_{1}}a_{2,j_{2}-i_{1}}}\right) x^{j_{2}}=\sum_{j_{1}=0}^{\infty
}A_{2,j_{2}}x^{j_{2}},
\end{eqnarray}%
\begin{equation}
A_{2,j_{2}}=\sum_{i_{1}=0}^{j_{2}}{a_{1,i_{1}}a_{2,j_{2}-i_{1}}},
\end{equation}%
\begin{eqnarray}
\hspace{-1cm}&&f_{1}\circ f_{2}\circ f_{3} =\left( \sum_{i_{1}=0}^{\infty
}a_{1,i_{1}}x^{i_{1}}\right) \left( \sum_{i_{2}=0}^{\infty
}a_{2,i_{2}}x^{i_{2}}\right) \times \notag \\
\hspace{-1cm}&&
\left( \sum_{i_{3}=0}^{\infty
}a_{3,i_{3}}x^{i_{3}}\right) =  \sum_{i_{1}=0}^{\infty }a_{1,i_{1}}x^{i_{1}}\left(
\sum_{i_{2},i_{3}=0}^{\infty }a_{2,i_{2}}a_{3,i_{3}}x^{i_{2}+i_{3}}\right)
\nonumber\\
\hspace{-1cm}&&=\sum_{i_{1}=0}^{\infty }a_{1,i_{1}}x^{i_{1}}\left[ \sum_{j_{2}=0}^{\infty
}\left( \sum_{i_{2}=0}^{j_{2}}a_{2,i_{2}}a_{3,j_{2}-i_{2}}\right) x^{j_{2}}%
\right] =  \notag \\
\hspace{-1cm}&&\sum_{j_{3}=0}^{\infty }\left[ \sum_{i_{1}=0}^{j_{3}}%
\sum_{i_{2}=0}^{j_{3}-i_{1}}a_{1,i_{1}}a_{2,i_{2}}a_{3,j_{3}-i_{1}-i_{2}}%
\right] x^{j_{3}}=\notag \\
\hspace{-1cm}&&
\sum_{j_{3}=0}A_{3,j_{3}}x^{j_{3}},
\end{eqnarray}%
\begin{equation}
A_{3,j_{3}}=\sum_{i_{1}=0}^{j_{3}}%
\sum_{i_{2}=0}^{j_{3}-i_{1}}a_{1,i_{1}}a_{2,i_{2}}a_{3,j_{3}-i_{1}-i_{2}},
\end{equation}%
\begin{equation*}
.......,
\end{equation*}%
\begin{equation}
f_{1}\circ f_{2}\circ ...\circ f_{s}=\sum_{j_{s}=0}^{\infty }A_{s,j_{s}}{%
x^{j_{s}}},  \label{gens}
\end{equation}%
\bea \label{gens1}
A_{s,j_{s}}&=&\sum_{i_{1}=0}^{j_{s}}\sum_{i_{2}=0}^{j_{s}-i_{1}}...\times \nonumber\\
&&\sum_{i_{s-1}=0}^{j_{s}-i_{1}-...-i_{s-1}}{%
a_{1,i_{1}}a_{2,i_{2}}...a_{s,j_{s}-i_{1}-...-i_{s-1}}}. \nonumber\\
\eea

\section{Exact series solution of the relativistic mass equation for a
linear barotropic fluid}\label{sect2}

As a first example of an exact power series solution of the relativistic
mass Eq.~(\ref{meq}) we consider the case of the linear barotropic equation
of state $p=\gamma \rho c^{2}$. Using Eq.~(\ref{aa}), we rewrite the
equation of state (\ref{0}) in the form
\begin{equation}
P\left( \eta \right) =\gamma \epsilon \left( \eta \right) ,\gamma =\mathrm{%
constant},\gamma \in \left[ 0,1\right] .  \label{EOS}
\end{equation}%
Then the TOV Eq.~(\ref{tovad}) becomes
\begin{equation}
\frac{d\epsilon \left( \eta \right) }{d\eta }=-a\frac{\gamma +1}{\gamma }%
\frac{\epsilon \left( \eta \right) \left[ \gamma \epsilon \left( \eta
\right) \eta ^{3}+m\left( \eta \right) \right] }{\eta ^{2}\left[ 1-2am\left(
\eta \right) /\eta \right] }.  \label{t3}
\end{equation}

\subsection{Exact power series solution of the relativistic mass equation}

By using the linear barotropic equation of state the relativistic mass Eq.~(%
\ref{meq}) takes the form
\bea
&&\eta \left( 1-\frac{2am}{\eta }\right) \frac{d^{2}m}{d\eta ^{2}}+\left[
a\left( \frac{1}{\gamma }+5\right) \frac{m}{\eta }-2\right] \frac{dm}{d\eta }%
+\nonumber\\
&&a\left( 1+\gamma \right) \left( \frac{dm}{d\eta }\right) ^{2}=0,  \label{k1}
\eea%
or, equivalently,
\begin{equation}
\eta \frac{d^{2}m}{d\eta ^{2}}-2\frac{dm}{d\eta }-2am\frac{d^{2}m}{d\eta ^{2}%
}+\alpha \frac{m}{\eta }\frac{dm}{d\eta }+\beta \left( \frac{dm}{d\eta }%
\right) ^{2}=0,  \label{A2}
\end{equation}%
where for simplicity we have introduced the coefficients $\alpha $ and $%
\beta $ defined as $\alpha =a\left( 1/\gamma +5\right) $ and $\beta =a\left(
1+\gamma \right) $, respectively.

Eqs.~(\ref{k1}) or (\ref{A2}) must be solved with the initial conditions $%
m(0)=0$, and $\left( dm/d\eta \right) |_{\eta =0}=0$. Note that Eqs.~(\ref%
{k1}) and (\ref{A2}) are not in the autonomous form, that is, the
coefficients of the derivative $dm/d\eta $ depend both on the mass function $%
m\left( \eta \right) $ and the dimensionless radius $\eta $. \ In order to
solve Eq. (\ref{A2}) we will look for exact power series solution of the
equation. Therefore we can state the following

\textbf{Theorem 1}. \textit{The relativistic mass equation (\ref{A2})
describing the interior of a star with matter content described by a linear
barotropic equation of state }$P\left( \eta \right) =\gamma \epsilon \left(
\eta \right) $\textit{, $\gamma =\mathrm{constant}$, has an exact
non-singular convergent power series solution of the form}{%
\begin{equation}
m(\eta )=\sum_{n=1}^{\infty }c_{2n+1}\eta ^{2n+1},\eta \leq 1.  \label{ms}
\end{equation}%
}\textit{\ with the coefficients $c_{2n+1}$ obtained from the recursive
relation}
\begin{eqnarray}
c_{2n+1} &=&-\frac{a}{2\left( n-1\right) \left( 2n+1\right) \gamma }\times
\notag \\
&&\sum_{i=1}^{n-1}\left( 2n-2i+1\right) \Bigg[2\gamma \left( \gamma
+3\right) i-4\gamma n+\nonumber\\
&&\gamma ^{2}+6\gamma +1\Bigg]c_{2i+1}c_{2n-2i+1},n\geq
2.  \label{T}
\end{eqnarray}

\textbf{Proof.} In the following we will look for a convergent power series
solution of Eq.~({\ref{k1}), by choosing $m(\eta )$ in the form given by Eq.
}~({\ref{ms}). Then it is easy to show the relations $\frac{dm}{d\eta }%
=\sum_{n=1}^{\infty }\left( 2n+1\right) c_{2n+1}\eta ^{2n}$, and $\frac{%
d^{2}m}{d\eta ^{2}}=2\sum_{n=1}^{\infty }n\left( 2n+1\right) c_{2n+1}\eta
^{2n-1}$, respectively. For the product of two power series we will use the
Cauchy convolution, so that
\bea
\left( \sum_{i=0}^{\infty }a_{i}\eta ^{i}\right) \left( \sum_{j=0}^{\infty
}b_{j}\eta ^{j}\right)& =&\sum_{i,j=0}^{\infty }a_{i}b_{j}\eta
^{i+j}=\nonumber\\
&&\sum_{n=0}^{\infty }\left( \sum_{i=0}^{n}a_{i}b_{n-i}\right) \eta
^{n}.
\eea
Thus,
\begin{eqnarray}
\hspace{-0.7cm}&&\left( \frac{dm}{d\eta }\right) ^{2} =\sum_{i,j=1}^{\infty }\left(
2i+1\right) \left( 2j+1\right) c_{2i+1}c_{2j+1}\eta ^{2i+2j} = \notag \\
\hspace{-0.7cm}&&\sum_{n=1}^{\infty }\Bigg[\sum_{i=1}^{n}\left( 2i+1\right) \left(
2n-2i+1\right) c_{2i+1}c_{2n-2i+1}\Bigg]\eta ^{2n},
\end{eqnarray}%
\bea
&&\frac{m}{\eta }\frac{dm}{d\eta }=\sum_{i,j=1}^{\infty }c_{2i+1}\left(
2j+1\right) c_{2j+1}\eta ^{2i+2j}=\nonumber\\
&&\sum_{n=1}^{\infty }\left[
\sum_{i=1}^{n}\left( 2n-2i+1\right) c_{2i+1}c_{2n-2i+1}\right] \eta ^{2n},
\eea
\bea
\hspace{-0.6cm}&&m\frac{d^{2}m}{d\eta ^{2}}=\sum_{i,j=1}^{\infty }c_{2i+1}2j\left(
2j+1\right) c_{2j+1}\eta ^{2i+2j}=\nonumber\\
\hspace{-0.6cm}&&\sum_{n=1}^{\infty }\left[
\sum_{i=1}^{n}2\left( n-i\right) \left( 2n-2i+1\right) c_{2i+1}c_{2n-2i+1}%
\right] \eta ^{2n}.
\eea

Hence by substituting these results into Eq.~(\ref{A2}) gives immediately
\begin{eqnarray}
&&\sum_{n=1}^{\infty }\Bigg\{2\left( n-1\right) \left( 2n+1\right)
c_{2n+1}+\nonumber\\
&&\sum_{i=1}^{n}\Bigg[-4a\left( n-i\right) \left( 2n-2i+1\right)
c_{2i+1}c_{2n-2i+1}+  \notag \\
&&\alpha \left( 2n-2i+1\right) c_{2i+1}c_{2n-2i+1}+\beta \left( 2i+1\right)\times \nonumber\\
&&\left( 2n-2i+1\right) c_{2i+1}c_{2n-2i+1}\Bigg]\Bigg\}\eta ^{2n}=0,
\end{eqnarray}
where we have transformed all the products of the power series by using the
Cauchy convolution. After using the definitions of $\alpha $ and $\beta $,
we obtain
\begin{eqnarray}
\hspace{-0.7cm}&&\sum_{n=1}^{\infty }\Bigg\{ 2\left( n-1\right) \left( 2n+1\right) \gamma
c_{2n+1}+a\sum_{i=1}^{n}\left( 2n-2i+1\right) \nonumber\\
\hspace{-0.7cm}&&\left[ 2\gamma \left( \gamma
+3\right) i+\gamma ^{2}+2\gamma \left( 3-2n\right) +1\right]c_{2i+1}c_{2n-2i+1}\Bigg\}\times \nonumber\\
\hspace{-0.7cm}&& \eta ^{2n}=0,n\geq 2.
\end{eqnarray}

By solving the above equation for the coefficients $c_{2n+1}$ gives the
recursive relation (\ref{T}) for the coefficients of the series
representation of the mass function. This ends the proof of \textbf{Theorem 1%
}.

For the values of the coefficients $c_{2n+1}$ we obtain the following
explicit expressions
\begin{equation}
c_{5}=-\frac{3a(\gamma +1)(3\gamma +1)}{10\gamma }c_{3}^{2},  \label{5}
\end{equation}%
\begin{equation}
c_{7}=\frac{3a^{2}(\gamma +1)(3\gamma +1)\left( 15\gamma ^{2}+9\gamma
+4\right) }{140\gamma ^{2}}c_{3}^{3},  \label{7}
\end{equation}%
\bea\label{9}
c_{9}&=&-\frac{a^{3}(\gamma +1)(3\gamma +1)}{2520\gamma ^{3}}\Bigg( 945\gamma ^{4}+864\gamma
^{3}+618\gamma ^{2}+\nonumber\\
&&200\gamma +61\Bigg) c_{3}^{4},
\eea
\bea\label{11}
c_{11}&=&\frac{a^{4}(\gamma +1)(3\gamma +1)}{184800\gamma ^{4}}\Bigg( %
85050\gamma ^{6}+91665\gamma ^{5}+\nonumber\\
&&80892\gamma ^{4}+38832\gamma ^{3}+17936\gamma ^{2}+4239\gamma +\nonumber\\
&&1258\Bigg) c_{3}^{5},
\eea
\begin{eqnarray} \label{14}
c_{13} &=&-\frac{a^{5}(\gamma +1)(3\gamma +1)}{12012000\gamma ^{5}}\Bigg(%
7016625\gamma ^{8}+\nonumber\\
&&8057475\gamma ^{7}+7978905\gamma ^{6}+4456683\gamma ^{5}+\nonumber\\
&&2486451\gamma ^{4}+
839697\gamma ^{3}+346075\gamma ^{2}+\nonumber\\
&&61953\gamma +22952\Bigg)c_{3}^{6},
\end{eqnarray}%
\begin{eqnarray}
c_{15} &=&\frac{a^{6}(\gamma +1)(3\gamma +1)}{5045040000\gamma ^{6}}\Bigg(%
3831077250\gamma ^{10}+\nonumber\\
&&4428596025\gamma ^{9}+4702427055\gamma
^{8}+\nonumber\\
&&2757559491\gamma ^{7}+ 1705375683\gamma ^{6}+\nonumber\\
&&636216069\gamma ^{5}+311382965\gamma
^{4}+72456873\gamma ^{3}+\nonumber\\
&&36302375\gamma ^{2}+3752022\gamma + 2703152\Bigg)c_{3}^{7},  \label{15}
\end{eqnarray}%
\begin{equation*}
.......  \label{16}
\end{equation*}

Using Eqs.~(\ref{mcad}) and (\ref{ms}), we obtain the energy density of the
matter inside a general relativistic star described by a linear barotropic
equation of state as
\begin{equation}
\epsilon \left( \eta \right) =\sum_{n=1}^{\infty }{\left( 2n+1\right)
c_{2n+1}\eta ^{2n-2}}.  \label{E}
\end{equation}%
By estimating the energy density at the center of the star $\eta =0$ gives $%
\epsilon \left( 0\right) =1=3c_{3}$, which fixes the value of the constant $%
c_{3}$ as
\begin{equation}
c_{3}=\frac{1}{3}.  \label{c3}
\end{equation}%
By inserting Eq.~(\ref{ms}) into Eq.~(\ref{lambda}), we obtain the metric
tensor component $e^{-\lambda }$ as
\begin{equation}
e^{-\lambda \left( \eta \right) }=1-2a\frac{m\left( \eta \right) }{\eta }%
=1-2a\sum_{n=1}^{\infty }{c_{2n+1}\eta ^{2n}}.
\end{equation}%
By substituting Eq.~(\ref{EOS}) into Eq.~(\ref{f4}), then the latter
equation can be integrated to yield%
\begin{equation}
e^{\nu \left( \eta \right) }=e^{\nu (0)}\left[\frac{\epsilon (\eta)}{\epsilon (0)}\right]^{-\frac{2\gamma }{1+\gamma }}%
,  \label{AB}
\end{equation}%
where $e^{\nu (0)}$ is the value of the metric coefficient at the center of the star, and $\epsilon (0)=1$. With the help of Eq.
(\ref{E}) we rewrite Eq.~(\ref{AB}) in the form
\begin{equation}
e^{\nu \left( \eta \right) }=e^{\nu (0)}\left[ \sum_{n=1}^{\infty }{\left( 2n+1\right)
c_{2n+1}\eta ^{2n-2}}\right] ^{-\frac{2\gamma }{1+\gamma }},\gamma \neq -1.
\end{equation}%
 Thus the
interior line element for a fluid sphere satisfying a linear barotropic
equation of state takes the form
\bea
ds^{2}&=&c^{2}e^{\nu (0)}\left[ \sum_{n=1}^{\infty }{\left( 2n+1\right) c_{2n+1}\eta
^{2n-2}}\right] ^{-\frac{2\gamma }{1+\gamma }}dt^{2}-\nonumber\\
&&\frac{1}{%
1-2a\sum_{n=1}^{\infty }{c_{2n+1}\eta ^{2n}}}dr^{2}-r^{2}d\Omega ^{2},\gamma
\neq -1.\nonumber\\
\eea

At the surface of the barotropic matter distribution $\epsilon (1)=\epsilon
_{S}=\rho _{S}/\rho _{c}=\mathrm{constant}$, where $\rho _{S}=\rho (R) $ is
the density of the barotropic fluid distribution on the boundary separating
the two phases. Thus we obtain $e^{\nu (R)}=e^{\nu (0)}\left( \rho _{S}/\rho
_{c}\right) ^{-2\gamma /(1+\gamma )}$. Moreover, it follows that on the boundary
$\eta =1$ of the barotropic component the coefficients $c_{2n+1} $ must
satisfy the condition $\sum_{n=1}^{\infty }\left( 2n+1\right)
c_{2n+1}=\epsilon _{S}$.

For the $e^{-\lambda }$ metric tensor component at the star surface we obtain $e^{-\lambda
}=1-2GM_{S}/c^{2}R$.

In order to test the accuracy of our power series solution we consider the
cases $\gamma =1/3$ and $\gamma =1$, respectively, corresponding to the
radiation fluid ($\gamma =1/3$), and stiff fluid ($\gamma =1$) equations of
state, respectively. The comparisons between the series solution of the TOV
and continuity equations, obtained via the solution of the relativistic mass
equation, and the exact numerical solution, computed by numerically
integrating the coupled system of Eqs.~(\ref{tovad}) and (%
\ref{mcad}) is represented in Fig.~\ref{fig1}.

\begin{figure*}[tbp]
\includegraphics[width=7.5cm, angle=0]{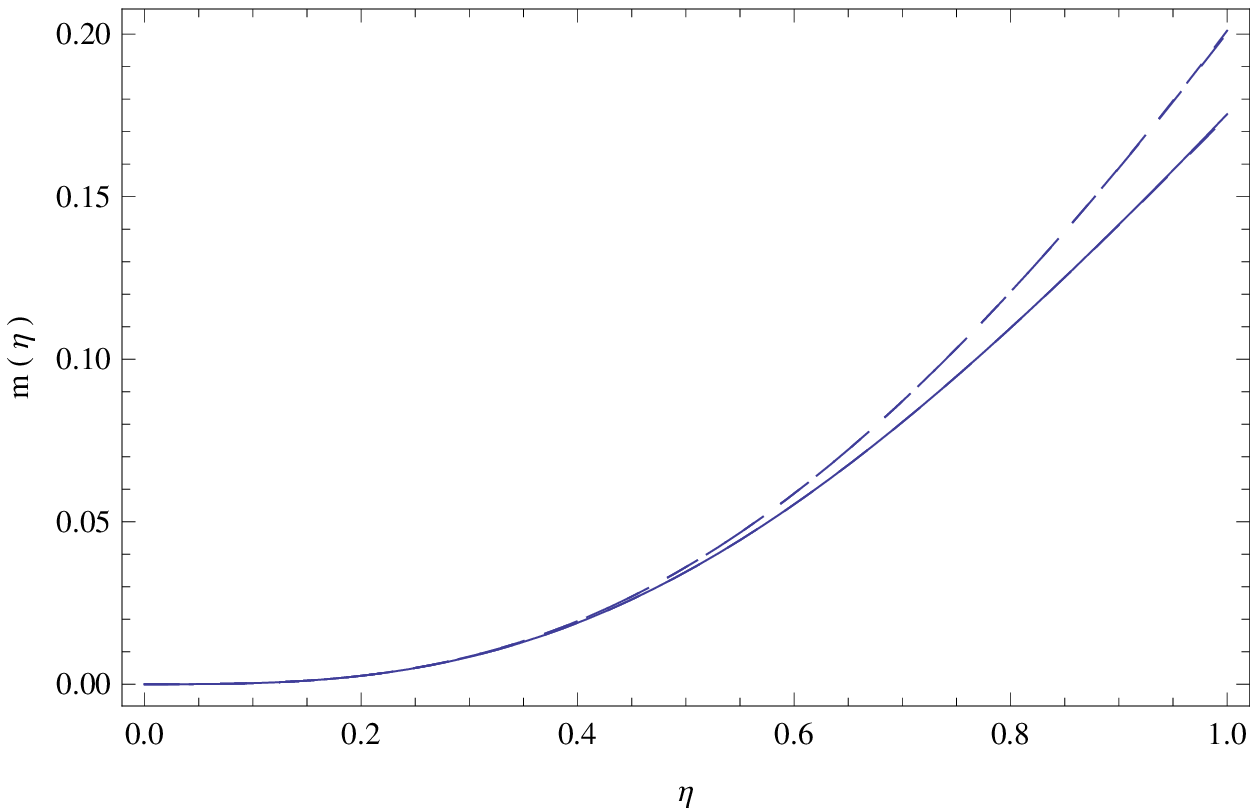} %
\includegraphics[width=7.5cm, angle=0]{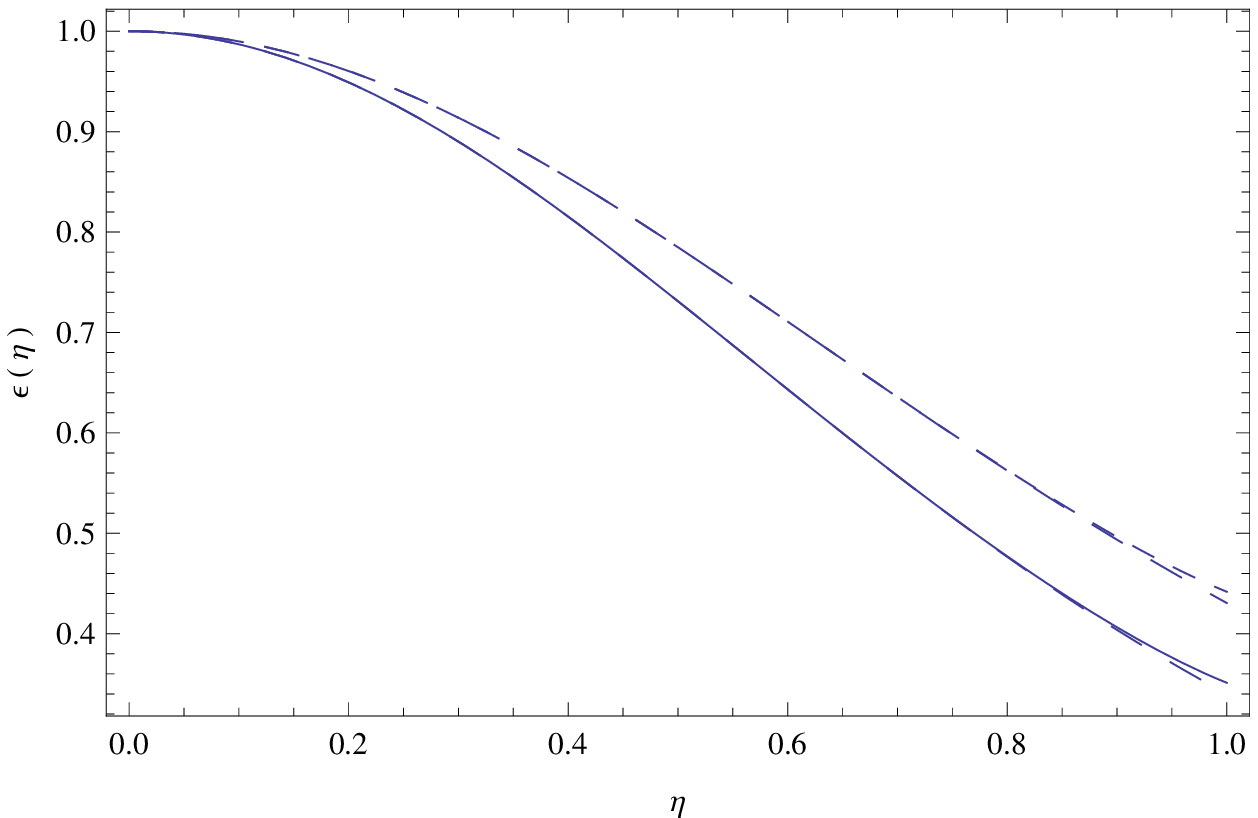}
\caption{Variation of the dimensionless mass $m(\eta)$ (left figure) and dimensionless energy density $\epsilon (\eta)$ for the radiation fluid star, with $\gamma =1/3$,  and for the stiff fluid equation of
state star, with $\gamma =1$, respectively. The dashed curve represents the
numerical solution of the TOV and continuity Eqs.~(\ref{tovad}) and (%
\ref{mcad}), while the solid and dotted curves represent the power series
solution of the relativistic mass equation, truncated to $n=7$. For the
radiation fluid star the surface density is fixed at $\epsilon %
_S=0.351165$, corresponding to $a=0.9876$. For the stiff fluid star $%
\epsilon _S=0.441812$, and $a=0.76389$. }
\label{fig1}
\end{figure*}

To numerically integrate Eqs.~(\ref{tovad}) and (%
\ref{mcad}) we have used the NDSolve command of the Mathematica software \citep{Wolfram},  which finds solutions iteratively, and by using the default setting of Automatic for AccuracyGoal and PrecisionGoal. The power
series solution has been truncated to seven terms only. Overall, even with
this small number of terms, the power series solution gives a good
approximation of the exact solution obtained by numerical integration of the
structure equations of the linear barotropic relativistic star.

For the radiation fluid star we have adopted the values $a=0.9876$, giving a
surface density $\epsilon _{S}=\epsilon (1)=0.351165$, with the total
dimensionless mass obtained as $m(1)=0.175377$. The physical parameters of
this stellar model are given by
\begin{equation}
R=10.2978\times \left( \frac{\rho _{c}}{10^{15}\;\mathrm{g/cm^{3}}}\right)
^{-1/2}\;\mathrm{km},
\ee
\be
M_{S}=1.2033\times \left( \frac{\rho _{c}}{10^{15}\;%
\mathrm{g/cm^{3}}}\right) ^{-1/2}\;M_{\odot }.
\end{equation}

For the stiff fluid star, with $\gamma =1$, $a=0.76389$, giving a surface
density of $\epsilon _{S}=0.441812$, and a total dimensionless mass of $%
m(1)=0.201045$. The global parameters of this high density star model can be
obtained as
\begin{equation}
R=9.05665\times \left( \frac{\rho _{c}}{10^{15}\;\mathrm{g/cm^{3}}}\right)
^{-1/2}\;\mathrm{km},
\ee
\be
M_{S}=0.93838\times \left( \frac{\rho _{c}}{10^{15}\;%
\mathrm{g/cm^{3}}}\right) ^{-1/2}\;M_{\odot }.
\end{equation}


\subsection{Matching with a constant density atmosphere}

Now we match the interior metric of the fluid sphere with matter content satisfying a linear barotropic equation of state to
the metric corresponding to a constant density atmosphere, with matter
density $\rho =\rho _{S}=\mathrm{constant}$, and pressure $p_{c}(r)$,
respectively. This metric is matched on the star's surface with the exterior
Schwarzschild metric, given by
\bea
ds^{2}&=&c^{2}\left( 1-\frac{2GM_{tot}}{c^{2}R_{tot}}\right) dt^{2}-\nonumber\\
&&\frac{1}{%
1-2GM_{tot}/c^{2}R_{tot}}dr^{2}-r^{2}d\Omega ^{2},  \label{10a}
\eea
where $M_{tot}=M_S+M_c$ and $R_{tot}=R+R_c$ are the total mass and radius of the star,
including both the linear barotropic and the constant density components. We
assume that the metric functions $g_{tt}$, $g_{rr}$ and $\partial
g_{tt}/\partial r$ are all continuous at both the contact region between the
barotropic and constant density matter, as well as at the vacuum boundary
surface of the star. In the constant density region we obtain first
\bea
m_{c}(r)&=&4\pi \rho _{S}\int_{R}^{r}{r^{2}dr}=\frac{4\pi \rho _{S}}{3}\left(
r^{3}-R^{3}\right) ,\nonumber\\
&&R\leq r\leq R_{tot},
\eea
\bea\label{10b}
e^{-\lambda (r)}&=&1-\frac{2G\left[M_S+4\pi \rho _S\left(r^3-R^3\right)/3\right]}{c^2r},\nonumber\\
&&R\leq r\leq R_{tot}.
\eea
The continuity of $\lambda $ at $r=R_{tot}$, $e^{-\lambda
_{c}(R)}=1-2GM_{tot}/c^{2}R_{tot}$ fixes the value of the surface density of the
linear barotropic region as
\begin{equation}
\rho _{S}=\frac{3\left(M_{tot}-M_S\right)}{4\pi\left(R_{tot}^3- R^{3}\right)}.
\end{equation}%
For the total mass of the star $M_{tot}$ from Eqs.~(\ref{10a}) and (\ref{10b}%
) we obtain
\be\label{45}
M_{tot}=M_S+\frac{4\pi}{3}\rho _S\left(R_{tot}^3-R^3\right).
\ee

In the constant density region Eq.~(\ref{f4}) can be integrated to give
\begin{equation}
e^{\nu _{c}\left( r\right) }=\frac{C}{\left[ \rho _{S}c^{2}+p_{c}\left(
r\right) \right] ^{2}},R\leq r\leq R_{tot}.
\end{equation}%
For $r=R$ we have $p_c(R)=\gamma \rho _Sc^2$,  $e^{\nu _c (R)}=1-2GM_S/c^2R$, giving for the integration constant $C$ the value  $C=\left(1-2GM_S/c^2R\right)\left(1+\gamma \right)^2\left(\rho _Sc^2\right)^2$, respectively. Thus we obtain
\be
e^{\nu _{c}\left( r\right) }=\left(1-\frac{2GM_S}{c^2R}\right)\frac{\left(1+\gamma \right)^2\left(\rho _Sc^2\right)^2}{\left[ \rho _{S}c^{2}+p_{c}\left(
r\right) \right] ^{2}},R\leq r\leq R_{tot}.
\ee

 On the surface of the star $p_{c}\left(
R_{tot}\right) =0$, and therefore
\begin{equation}\label{42}
\frac{2GM_{tot}}{c^{2}R_{tot}}=1-\left(1+\gamma \right)^2\left(1-\frac{2GM_S}{c^2R}\right).
\end{equation}%
Eq.~(\ref{42}) gives the  total mass-total radius ratio of the star, once the mass, radius and equation of state of the core described by a linear barotropic equation of state are known.

In the next Section, we shall consider power series solutions of the
relativistic mass equation for polytropic fluids.

\section{Exact power series solutions of the relativistic mass equation for
polytropic stars}\label{sect3}

Polytropic models play an important role in the galactic dynamics and in the
theory of stellar configuration and evolution \citep{1s}. In particular,
polytropic models with $n=1$ can be used to model Bose-Einstein Condensate
dark matter \citep{BoHa07}, and Bose-Einstein Condensate stars \citep{ChHa},
respectively. For a polytropic system, the interior structure of the compact
objects can be described by an equation of state of the form
\begin{equation}
p\left( r\right) =K\rho ^{1+\frac{1}{n}}\left( r\right) ,  \label{p1}
\end{equation}%
where $p\left( r\right) $ and $\rho \left( r\right) $ are the pressure and
the energy density respectively, while $K$ and $n$ are constants. The
constant $n$ is called the polytropic index. In galactic dynamics $n>1/2$,
and no polytropic stellar system can be homogenous \citep{2s}. In the case of
the theory of stellar structure and evolution, in general, $n$ ranges from $%
0 $ to $\infty $ \citep{1s,3s}. Similarly to the previous Section, with the
help of Eqs.~(\ref{p1}) and (\ref{aa}), we obtain the polytropic equation of
state in a dimensionless form given by
\begin{equation}
P(\eta )=k\epsilon ^{1+1/n}(\eta ),  \label{af}
\end{equation}%
where we have denoted the constant $k$ as $k=K\rho
_{c}^{1/n}/c^{2} $.  By inserting Eq.~(\ref{af}) into Eq.~(\ref{f4}), then the
latter can be integrated to give
\begin{equation}
e^{\nu (\eta )}=e^{\nu (0)}\left[\frac{ 1+k\epsilon ^{1/n}(\eta )}{1+k}\right] ^{-2(1+n)}.
\label{nupol}
\end{equation}%
 On the surface of the polytropic star,
corresponding to $\eta =1$, the metric tensor coefficient (\ref{nupol}) must
be matched with the Schwarzschild line element, thus giving
\begin{equation}
e^{\nu (1)}=e^{\nu (0)}\left[\frac{ 1+k\epsilon ^{1/n}(\eta )}{1+k}\right] ^{-2(1+n)}=1-\frac{2GM_{S}}{%
c^{2}R},
\end{equation}%
where $M_{S}$ and $R$ are the mass and the radius of the star, respectively.
A vanishing surface energy density $\epsilon (1)=0$ would give, for $n>0$, the value of the metric tensor coefficient at the center of the star as
\be
e^{\nu (0)}=\frac{1}{\left(1+k\right)^{2(1+n)}}\left(1-\frac{2GM_{S}}{c^{2}R}\right).
\ee
 By  assuming that at the surface of polytropic star the density
(and the pressure) does not vanish, and that $\epsilon (1)=\epsilon _{S}\neq 0$,
the surface density is determined by the equation
\begin{equation}
\epsilon _{S}^{1/n}= \frac{1}{k}\left\{ \left[\frac{(1+k)}{e^{\nu (0)}}\left( 1-\frac{2GM_{S}}{c^{2}R}%
\right)\right] ^{-\frac{1}{2(1+n)}}-1\right\}.
\end{equation}%

By setting again $M^{\ast }$ and $a$ as $M^{\ast }=4\pi \rho _{c}R^{3}$ $\ $%
and $a=4\pi G\rho _{c}R^{2}/c^{2}$, respectively, and with the help of Eq.~(%
\ref{af}), the TOV Eq.~(\ref{tovad}) gives a differential equation for $%
\epsilon (\eta )$
\bea
\frac{d\epsilon }{d\eta }&=&-\frac{an\epsilon
(\eta )}{k\left( n+1\right) }\times \nonumber\\
&&\frac{\left[ \epsilon ^{-1/n}(\eta )+k\right] \left[ k\eta ^{3}\epsilon
^{1+1/n}(\eta )+m(\eta )\right] }{\eta ^{2}\left[ 1-2am(\eta )/\eta \right] }%
.  \label{nm}
\eea%
By inserting Eq.~(\ref{mcad}) into Eq.~(\ref{nm}), the latter gives the
differential equation for the relativistic mass function $m(\eta )$ as
\begin{eqnarray}
&&\eta ^{2}\left[ 1-2a\frac{m(\eta )}{\eta }\right] \frac{d^{2}m}{d\eta ^{2}}%
+\frac{an}{n+1}\eta \Bigg[ k\left( \frac{1}{\eta ^{2}}\frac{dm}{d\eta }%
\right) ^{\frac{1}{n}}+\nonumber\\
&&1\Bigg] \left( \frac{dm}{d\eta }\right) ^{2}+
\Bigg\{ \frac{a}{k(n+1)}\Bigg[ k(5n+4)+\nonumber\\
&&n\left( \frac{1}{\eta ^{2}}\frac{dm%
}{d\eta }\right) ^{-\frac{1}{n}}\Bigg] m(\eta )-2\eta \Bigg\} \frac{dm}{%
d\eta }=0.  \label{kk}
\end{eqnarray}%
In the following, we consider first that $n=1$, and we show that for this
case a power series solution of the relativistic mass equation does exist,
by explicitly constructing it. As a next step in our study we will consider
the power series solution of the relativistic mass equation for arbitrary $n$%
.

\subsection{The case $n=1$}

For $n=1$, which corresponds to a polytropic equation of state of the form $%
p\propto \rho ^{2}$, Eq.~(\ref{kk}) takes the form
\bea
&&2\eta ^{2}\left[ 1-2a\frac{m(\eta )}{\eta }\right] \frac{d^{2}m}{d\eta ^{2}}+%
\left[ 9am(\eta )-4\eta \right] \frac{dm}{d\eta }+\nonumber\\
&&a\eta \left( \frac{dm}{%
d\eta }\right) ^{2}+\frac{ak}{\eta }\left( \frac{dm}{d\eta }\right) ^{3}+%
\frac{a}{k}\eta ^{2}m(\eta )=0.  \label{se}
\eea
For mathematical convenience, we rewrite Eq.~(\ref{se}) in the form
\bea
&&\eta ^{2}\frac{d^{2}m}{d\eta ^{2}}-2\eta \frac{dm}{d\eta }+\frac{a}{2k}\eta
^{2}m-2a\eta m\frac{d^{2}m}{d\eta ^{2}}+\nonumber\\
&&\frac{9a}{2}m\frac{dm}{d\eta }+\frac{%
a}{2}\eta \left( \frac{dm}{d\eta }\right) ^{2}+\frac{ak}{2}\frac{1}{\eta }%
\left( \frac{dm}{d\eta }\right) ^{3}=0.  \label{w3}
\eea

We will look again for a power series solution of Eq. (\ref{se}), and
therefore our results can be summarized in the following

\textbf{Theorem 2}. \textit{The relativistic mass equation (\ref{se})
describing the interior physical and geometrical properties of a general
relativistic polytropic fluid sphere with polytropic index $n=1$ has an
exact non-singular power series solution $m(\eta )=\sum_{l=1}^{\infty
}c_{2l+1}\eta ^{2l+1}$, with coefficients $c_{2l+1}$, $l=1,2,..,\infty $
given by the recursive relation}
\begin{eqnarray}
c_{2l+3} &=&-\frac{a}{2l\left( 2l+3\right) }\Bigg[\frac{1}{2k}%
c_{2l+1}-\sum_{i=1}^{l}\left( 4l-5i-1\right) \times \nonumber\\
&&\left( 2l-2i+3\right)
c_{2i+1}c_{2l-2i+3}+\frac{k}{2} \sum_{i=1}^{l}\sum_{j=1}^{l-i+1} \notag  \label{eq65} \\
&&\left( 2i+1\right) \left( 2j+1\right)
\left( 2l-2i-2j+5\right) \times \nonumber\\
&&
c_{2i+1}c_{2j+1}c_{2l-2i-2j+5}\Bigg],l\in \lbrack
1,\infty ).
\end{eqnarray}

\textbf{Proof.} By inserting the power series representation of the mass,
given by Eq.~(\ref{ms}), into Eq. (\ref{w3}), the latter becomes
\begin{eqnarray}
\hspace{-0.5cm} &&2\sum_{l=1}^{\infty }(2l+1)\left( l-1\right) c_{2l+1}\eta ^{2l+1}+\frac{a%
}{2k}\sum_{l=1}^{\infty }c_{2l+1}\eta ^{2l+3}-\nonumber\\
\hspace{-0.5cm} &&4a\sum_{i,j=1}^{\infty
}j\left( 2j+1\right) c_{2i+1}c_{2j+1}\eta ^{2i+2j+1}+  \notag  \label{w5} \\
\hspace{-0.5cm} &&\frac{9a}{2}\sum_{i,j=1}^{\infty }\left( 2j+1\right) c_{2i+1}c_{2j+1}\eta
^{2i+2j+1}+\nonumber\\
\hspace{-0.5cm} &&\frac{a}{2}\sum_{i,j=1}^{\infty }\left( 2i+1\right) \left(
2j+1\right) c_{2i+1}c_{2j+1}\eta ^{2i+2j+1}+  \notag \\
\hspace{-0.5cm} &&\frac{ak}{2}\sum_{i,j,h=1}^{\infty }\left( 2i+1\right) \left( 2j+1\right)
\left( 2h+1\right) c_{2i+1}\times \nonumber\\
\hspace{-0.5cm} &&c_{2j+1}c_{2h+1}\eta ^{2i+2j+2h-1}=0.
\end{eqnarray}

In the first sum in Eq.~(\ref{w5}) the term corresponding to $l=1$
identically vanishes. Hence we can replace in the first sum $l$ by $l+1$.
In the terms containing $\eta ^{2i+2j+1}$ we use the Cauchy convolution of
the power series, and takes $2i+2j+1=2l+3$, or $i+j=l+1$.  The last term in
Eq. (\ref{w5}) can be transformed as follows
\begin{eqnarray}
\hspace{-0.8cm}&&\sum_{i,j,h=1}^{\infty }\left( 2i+1\right) \left( 2j+1\right) \left(
2h+1\right) c_{2i+1}c_{2j+1}c_{2h+1} \times \nonumber\\
\hspace{-0.8cm}&&\eta ^{2i+2j+2h-1}
= \sum_{i=1}^{\infty }\left( 2i+1\right) c_{2i+1}\eta ^{2i-1}\times \nonumber\\
\hspace{-0.8cm}&&\left[
\sum_{j,h=1}^{\infty }\left( 2j+1\right) \left( 2h+1\right)
c_{2j+1}c_{2h+1}\eta ^{2j+2h}\right]=  \notag \\
\hspace{-0.8cm} &&\sum_{i=1}^{\infty }\left( 2i+1\right) c_{2i+1}\eta
^{2i-1}\sum_{r=1}^{\infty }\Bigg[ \sum_{j=1}^{r}\left( 2j+1\right) \left(
2r-2j+1\right) \nonumber\\
\hspace{-0.8cm}&&c_{2j+1}c_{2r-2j+1}\Bigg] \eta ^{2r}=  \sum_{l=1}^{\infty }\Bigg[ \sum_{i=1}^{l}\sum_{j=1}^{l-i+1}\left(
2i+1\right) \left( 2j+1\right)\notag \\
\hspace{-0.8cm} &&
 \left( 2l-2i-2j+5\right)
c_{2i+1}c_{2j+1}c_{2l-2i-2j+5}\Bigg] \eta ^{2l+3}.
\end{eqnarray}%
Therefore it follows that the coefficients $c_{2l+1}$ must satisfy the
algebraic condition
\begin{eqnarray}
&&\sum_{l=1}^{\infty }\Bigg[2l\left( 2l+3\right) c_{2l+3}+\frac{a}{2k}%
c_{2l+1}-\nonumber\\
&&a\sum_{i=1}^{l}\left( 4l-5i-1\right) \left( 2l-2i+3\right)
c_{2i+1}c_{2l-2i+3}+  \notag \\
&&\frac{ak}{2}\sum_{i=1}^{l}\sum_{j=1}^{l-i+1}\left( 2i+1\right) \left(
2j+1\right) \left( 2l-2i-2j+5\right) \nonumber\\
&&c_{2i+1}c_{2j+1}c_{2l-2i-2j+5}\Bigg]%
\eta ^{2l+3}=0.
\end{eqnarray}

By solving the above equation for the coefficients $c_{2l+3}$ gives the
stated recursive relationship. This ends the proof of \textbf{Theorem 2}.

The condition $\epsilon (0)=\left. \sum_{l=1}^{\infty }(2l+1)c_{2l+1}\eta
^{2l-2}\right\vert _{\eta =0}=1$ fixes the coefficient $c_{3}$ in the series
expansion (\ref{ms}) as $c_{3}=1/3$. Then Eq.~(\ref{eq65}) gives for the
coefficients $c_{2l+1}$, $l=1,..,7$, the values
\begin{equation}
c_{5}=-\frac{a(k+1)(3k+1)}{60k},
\end{equation}%
\begin{equation}
c_{7}=\frac{a^{2}(k+1)(3k+1)\left( 45k^{2}-2k+3\right) }{10080k^{2}},
\label{c_7}
\end{equation}%
\bea
c_{9}&=&-\frac{a^{3}(k+1)(3k+1)}{%
362880k^{3}}\Bigg( 525k^{4}-62k^{3}+82k^{2}-\nonumber\\
&&30k+1\Bigg) ,
\eea
\bea
c_{11}&=&\frac{a^{4}(k+1)(3k+1)}{
63866880k^{4}}\Bigg(
33075k^{6}-6630k^{5}+\nonumber\\
&&8947k^{4}-3964k^{3}+1777k^{2}-206k+1\Bigg) ,
\eea
\begin{eqnarray}
c_{13} &=&-\frac{a^{5}(k+1)(3k+1)}{49816166400k^{5}}\Bigg(%
9823275k^{8}-2813850k^{7}+\nonumber\\
&&
3948156k^{6}-2032438k^{5}+  1202938k^{4}-\nonumber\\
&&
489670k^{3}+106892k^{2}-3178k+3\Bigg),
\end{eqnarray}%
\begin{eqnarray}
c_{15} &=&\frac{a^{6}(k+1)(3k+1)}{41845579776000k^{6}}\Bigg(%
3277699425k^{10}-\nonumber\\
&&1226770650k^{9}+1791820419k^{8}-  \notag \\
&&1054681792k^{7}+715276538k^{6}-\nonumber\\
&&389612036k^{5}+164352038k^{4}-47978464k^{3}+
\notag \\
&&3637621k^{2}-34210k+7\Bigg),
\end{eqnarray}%
\begin{equation*}
.......
\end{equation*}

It is interesting to compare the solution of the relativistic mass equation (\ref{w3}) with the exact solution of the {\it Newtonian} Lane-Emden equation (\ref{F1}), corresponding to $n=1$. By combining the non-relativistic hydrostatic equilibrium equation with the mass continuity equation we obtain
\be\label{LEN}
\frac{d}{dr}\left(\frac{r^2}{\rho}\frac{dp}{dr}\right)=-4\pi r^2 \rho.
\ee

By taking into account the polytropic equation of state with index $n=1$, $p=K\rho ^2$, Eq.~(\ref{LEN}) gives
\be\label{LEN1}
\frac{d}{dr}\left(r^2\frac{d\rho }{dr}\right)=-\frac{2\pi G}{K}r^2\rho.
\ee
With the help of the transformations $r=R\eta $, $\rho =\rho _c\epsilon $, $K=kc^2/\rho _c$, and by taking into account the definition of $a$, Eq.~(\ref{LEN1}) becomes
\be\label{LEN2}
\frac{d}{d\eta}\left(\eta ^2\frac{d\epsilon}{d\eta}\right)=-\frac{a}{2k}\eta ^2\epsilon.
\ee
Eq.~(\ref{LEN2})  has the non-singular solution
\be
\epsilon (\eta)=\left(\frac{2k}{a}\right)^{1/4}\frac{\sin\left[\left(a/2k\right)^{1/4}\eta\right]}{\eta},
\ee
satisfying the condition $\epsilon (0)=1$. The mass distribution of the Newtonian $n=1$ polytrope is given by
\bea
m(\eta )&=&\frac{\sqrt{2}}{\left( a/k\right) ^{3/4}}\Bigg\{ 2^{1/4}\sin \left[
\left( a/2k\right) ^{1/4}\eta \right] -\nonumber\\
&&\eta \left( a/k\right) ^{1/4}\cos %
\left[ \left( a/2k\right) ^{1/4}\eta \right] \Bigg\} .
\eea

The comparison of the exact numerical solution of the TOV and mass
continuity equations, and the power series solution of the relativistic mass
equations is presented in Fig.~\ref{fig3}.

\begin{figure*}[tbp]
\includegraphics[width=7.5cm, angle=0]{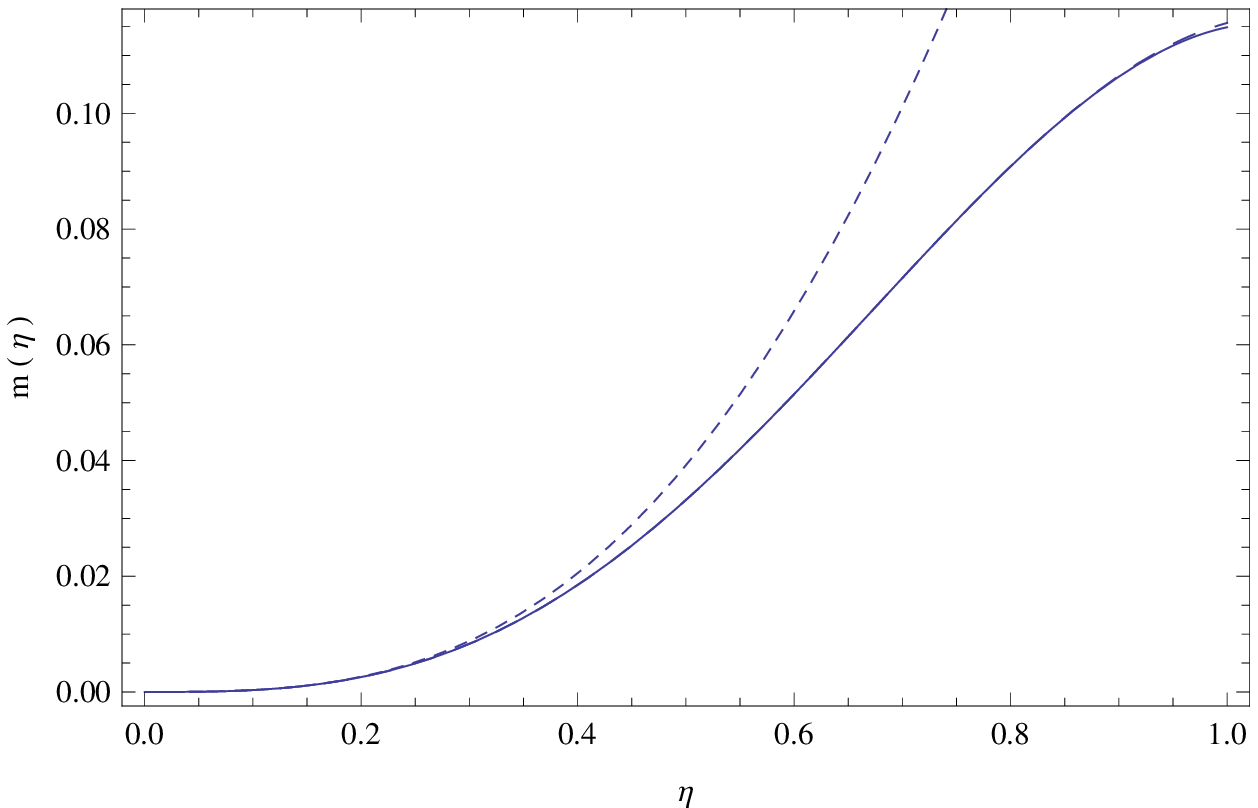} %
\includegraphics[width=7.5cm, angle=0]{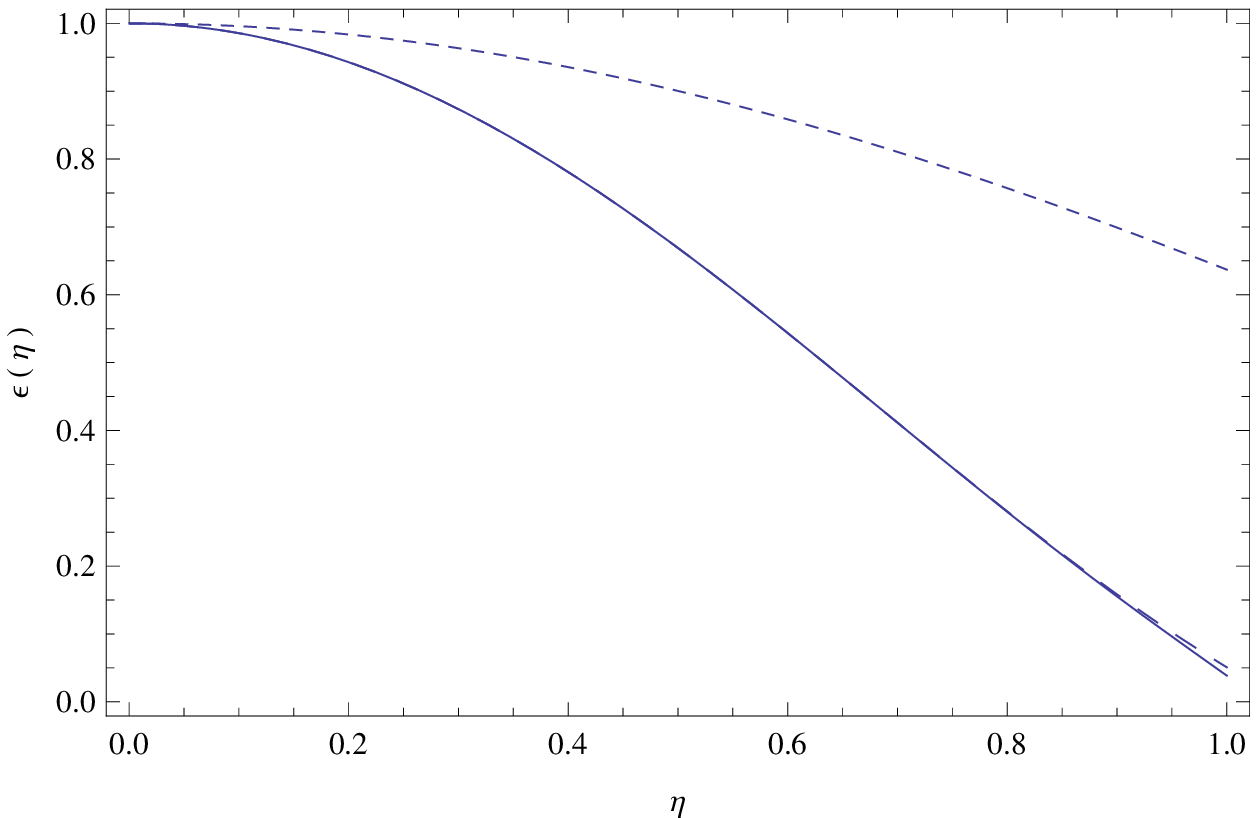}
\caption{Variation with respect to the dimensionless radial coordinate $%
\protect\eta $ of the dimensionless mass $m (\protect\eta)$ (left panel),
and of the dimensionless energy density $\protect\epsilon (\protect\eta )$
(right panel) for a polytropic star with index $n=1$ and $k=0.1$. The dashed
curves represent the numerical solution of the TOV and mass continuity Eqs.~(%
\protect\ref{mcad}) and (\protect\ref{t3}), while the solid curve represents
the power series solution of the relativistic mass equation, truncated to $%
l=7$. The surface density of the star is $\protect\epsilon _S=0.0385648$,
corresponding to $a=1.2145$. The dotted curve represents the Newtonian exact solution of the Lane-Emden equation for $n=1$ having the same numerical values of the parameters as the relativistic model. }
\label{fig3}
\end{figure*}

For the value of the
dimensionless parameter $k$ we have adopted the value $k=0.1$. As one can see from the Figures, there is an excellent agreement between the power series and the numerical solutions, respectively. The surface
energy density is fixed at $\epsilon _{S}=0.0385648$, which gives for the
coefficient $a$ the numerical value $a=1.2145$. The physical parameters of
the $n=1$ general relativistic star can be obtained as
\begin{equation}
R=11.4196\times \left( \frac{\rho _{c}}{10^{15}\;\mathrm{g/cm^{3}}}\right)
^{-1/2}\;\mathrm{km},
\ee
\be
M_{S}=1.07496\times \left( \frac{\rho _{c}}{10^{15}\;%
\mathrm{g/cm^{3}}}\right) ^{-1/2}\;M_{\odot }.
\end{equation}

On the other hand, as expected, the Newtonian non-relativistic solutions of the Lane-Emden equation give a poor description of the interior properties of dense general relativistic stars.

\section{The case of the arbitrary polytropic index $n$}

\label{sect4}

In order to obtain a convergent power series solution of Eq.~(\ref{kk}) for
arbitrary $n$ we will make use of the following

\textbf{Theorem 3 [Chang and Mott] \citep{CM}}. \textit{If $F(x)$ is an
arbitrary function of $x$, differentiable at $x=0$, and $A(z)=\sum_{j=1}^{%
\infty }{a_{j}z^{j}}$, then $G(z)=F\left[ A(z)\right] =\sum_{k=0}^{\infty }{%
g_{k}z^{k}}$, where $g_{k}=\sum_{h=0}^{k}{f_{h}\alpha _{hk}}$, where $%
f_{h}=\left. \frac{1}{h!}\frac{d^{h}}{dx^{h}}F(x)\right\vert _{x=0}$, and $%
\alpha _{hk}=\left. \frac{1}{k!}\frac{d^{k}}{dz^{k}}\left[ A(z)\right]
^{h}\right\vert _{z=0}$.}

\hspace{-0.5cm}\textbf{Proof.} By Taylor series expansion $F(x)=\sum _{h=0}^{\infty}{f_hx^h}
$, and $\left[A[z]\right]^h=\left(\sum _{j=1}^{\infty}{a_j z^j}%
\right)^h=\sum _{k=h}^{\infty} {\alpha _{hk}z^k}$, respectively. Then it immediately
follows
\bea
G(z)&=&F\left[A(z)\right]=\sum _{h=0}^{\infty}f_h \sum _{k=h}^{\infty}{\alpha
_{hk}z^k}=\nonumber\\
&&\sum _{k=0}^{\infty}\left(\sum _{h=0}^k{f_h\alpha _{hk}}%
\right)z^k=\sum _{k=0}^{\infty}g_kz^k.
\eea

With the use of \textbf{Theorem 3} we can now formulate the following

\textbf{Theorem 4}. \textit{The relativistic mass Eq.~(\ref{kk}), describing
the structure of polytropic general relativistic stars with arbitrary
polytropic index $n$, has an exact power series solution $m(\eta
)=\sum_{l=1}^{\infty }{c_{2l+1}\eta ^{2l+1}}$, with the coefficients of the
power series satisfying the recursive relation}
\begin{eqnarray}
\hspace{-0.5cm}&&c_{2l+3}=-\frac{a}{2l(2l+3)}\sum_{h=1}^{l}\Bigg\{\left( 2l-2h+3\right)\times \nonumber\\
 \hspace{-0.5cm}&&\left[ h\left( \frac{2(k+1)n}{n+1}+4\right) +\frac{(k+1)^{2}n}{k(n+1)}-4l%
\right] c_{2h+1}c_{2l-2h+3}+  \notag \\
\hspace{-0.5cm}&&\frac{n}{n+1}\sum_{i=1}^{l-h+1}(2l-2h-2i+3)\left[ (2i+1)k\alpha _{2h}^{+}+%
\frac{1}{k}\alpha _{2h}^{-}\right]\nonumber\\
\hspace{-0.5cm}&& c_{2i+1}c_{2l-2h-2i+3}\Bigg\},  \label{ff}
\end{eqnarray}%
\textit{where}
\bea
\alpha _{2j}^{\pm }&=&\frac{1}{\left( 2j\right) !}\left. \left[ \frac{d^{2j}}{%
d\eta ^{2j}}\left( \sum_{r=1}^{\infty }\left( 2r+1\right) c_{2r+1}\eta
^{2r-2}\right) ^{\pm 1/n}\right] \right\vert _{\eta =0},\nonumber\\
&&j=1,2,3,....
\eea

\textbf{Proof.} As a first step in our proof we rewrite Eq.~({\ref{kk}) in
the form
\begin{eqnarray}  \label{kk1}
&&\eta ^2\frac{d^2m}{d\eta ^2}-2\eta \frac{dm}{d\eta }-2a\eta m\frac{d^2m}{%
d\eta ^2}+\frac{an}{n+1}\eta \left(\frac{dm}{d\eta }\right)^2+\nonumber\\
&&\frac{a(5n+4)}{%
n+1}m\frac{dm}{d\eta}+
\frac{an}{n+1}k\eta \left(\frac{1}{\eta ^2}\frac{dm}{d\eta }%
\right)^{1/n}\left(\frac{dm}{d\eta }\right)^2+\nonumber\\
&&\frac{an}{k(n+1)}\left(\frac{1%
}{\eta ^2}\frac{dm}{d\eta }\right)^{-1/n}m\frac{dm}{d\eta }=0.
\end{eqnarray}
}

We will look again for a power series solution of Eq.~({\ref{kk1}), by
choosing $m(\eta )$ in the form $m(\eta )=\sum_{l=1}^{\infty }c_{2l+1}\eta
^{2l+1}$. Then we obtain immediately
\begin{eqnarray}
\hspace{-0.5cm}&&\left( \frac{1}{\eta ^{2}}\frac{dm}{d\eta }\right) ^{\pm \frac{1}{n}}
=\left( \sum_{r=1}^{\infty }\left( 2r+1\right) c_{2r+1}\eta ^{2r-2}\right)
^{\pm 1/n} = \notag \\
\hspace{-0.5cm}&&\sum_{j=0}^{\infty }\frac{1}{j!}\left. \left[ \frac{d^{j}}{d\eta ^{j}}%
\left( \sum_{r=1}^{\infty }\left( 2r+1\right) c_{2r+1}\eta ^{2r-2}\right)
^{\pm 1/n}\right] \right\vert _{\eta =0}\eta ^{j}.\nonumber\\
\end{eqnarray}%

By direct checking it can be shown that
\bea
&&\left. \left[ \frac{d^{j}}{d\eta ^{j}}\left( \sum_{r=1}^{\infty }\left(
2r+1\right) c_{2r+1}\eta ^{2r-2}\right) ^{\pm 1/n}\right] \right\vert _{\eta
=0}\equiv 0,\nonumber\\
&&j=1,3,5,....,
\eea
and thus
\begin{equation}
\left( \frac{1}{\eta ^{2}}\frac{dm}{d\eta }\right) ^{\pm \frac{1}{n}%
}=1+\sum_{j=1}^{\infty }\alpha _{2j}^{\pm }\eta ^{2j},
\end{equation}%
where
\bea
\hspace{-0.8cm}&&\alpha _{2j}^{\pm }=\frac{1}{\left( 2j\right) !}\left. \left[ \frac{d^{2j}}{%
d\eta ^{2j}}\left( \sum_{r=1}^{\infty }\left( 2r+1\right) c_{2r+1}\eta
^{2r-2}\right) ^{\pm 1/n}\right] \right\vert _{\eta =0},\nonumber\\
\hspace{-0.8cm}&&j=1,2,3....
\eea

Therefore it follows that
\begin{eqnarray}
\hspace{-1.5cm}&&\frac{an}{n+1}k\eta \left( \frac{1}{\eta ^{2}}\frac{dm}{d\eta }\right) ^{%
\frac{1}{n}}\left( \frac{dm}{d\eta }\right) ^{2} = \notag \\
 \hspace{-1.5cm}&&\frac{an}{n+1}k\eta \left( \frac{dm}{d\eta }\right) ^{2}+\frac{an}{n+1}%
k\sum_{h=1}^{\infty }\alpha _{2h}^{+}\eta ^{2h}\times\nonumber\\
\hspace{-1.5cm}&&\sum_{i,j=1}^{\infty }\left(
2i+1\right) \left( 2j+1\right) c_{2i+1}c_{2j+1}\eta ^{2i+2j+1} = \notag \\
\hspace{-1.5cm}&&\frac{an}{n+1}k\eta \left( \frac{dm}{d\eta }\right) ^{2}+\frac{an}{n+1}%
k\sum_{h=1}^{\infty }\alpha _{2h}^{+}\eta ^{2h}\nonumber\\
\hspace{-1.5cm}&&\sum_{r=1}^{\infty }\left[
\sum_{i=1}^{r}\left( 2i+1\right) \left( 2r-2i+1\right) c_{2i+1}c_{2r-2i+1}%
\right] \eta ^{2r+1} = \notag \\
\hspace{-1.5cm}&&\frac{an}{n+1}k\eta \left( \frac{dm}{d\eta }\right) ^{2}+\frac{an}{n+1}%
k\sum_{l=1}^{\infty }\Bigg[ \sum_{h=1}^{l}\sum_{i=1}^{l-h+1}\alpha
_{2h}^{+}\nonumber\\
\hspace{-1.5cm}&&\left( 2i+1\right) \left( 2l-2h-2i+3\right) c_{2i+1}c_{2l-2h-2i+3}%
\Bigg] \eta ^{2l+3},  \notag \\
\end{eqnarray}%
\begin{eqnarray}
&&\frac{an}{k(n+1)}\left( \frac{1}{\eta ^{2}}\frac{dm}{d\eta }\right)
^{-1/n}m\frac{dm}{d\eta } = \notag \\
 &&\frac{an}{k(n+1)}m\frac{dm}{d\eta }+\frac{an}{k(n+1)}\sum_{h=1}^{\infty
}\alpha _{2h}^{-}\eta ^{2h}\sum_{i,j=1}^{\infty }\left( 2j+1\right)\nonumber\\
&&c_{2i+1}c_{2j+1}\eta ^{2i+2j+1}
= \frac{an}{k(n+1)}m\frac{dm}{d\eta }+\frac{an}{k(n+1)}\nonumber\\
&&\sum_{h=1}^{\infty
}\alpha _{2h}^{-}\eta ^{2h}\sum_{r=1}^{\infty }\left[ \sum_{i=1}^{r}\left(
2r-2i+1\right) c_{2i+1}c_{2r-2i+1}\right] \eta ^{2r+1} = \notag \\
 &&\frac{an}{k(n+1)}m\frac{dm}{d\eta }+\frac{an}{k(n+1)}\sum_{l=1}^{\infty }%
\Bigg[ \sum_{h=1}^{l}\sum_{i=1}^{l-h+1}\alpha _{2h}^{-}\nonumber\\
&&\left(
2l-2h-2i+3\right) c_{2i+1}c_{2l-2h-2i+3}\Bigg] \eta ^{2l+3}.
\end{eqnarray}

Then we successively obtain
\begin{equation}
\hspace{0.5cm}\eta ^2\frac{d^2m}{d\eta ^2}-2\eta \frac{dm}{d\eta }=\sum
_{l=1}^{\infty}2l(2l+3)c_{2l+3}\eta ^{2l+3},
\end{equation}
\bea
\hspace{0.5cm}-2a\eta m\frac{d^2m}{d\eta ^2}&=&-4a \sum _{l=1}^{\infty}\Bigg[\sum
_{h=1}^l(l-h+1)(2l-\nonumber\\
&&\hspace{-0.1cm}2h+3)
c_{2h+1}c_{2l-2h+3}\Bigg]\eta ^{2l+3},
\eea
\bea
\hspace{0.5cm}\frac{an(k+1)}{n+1}\eta \left(\frac{dm}{d\eta }\right)^2&=&\frac{an(k+1)}{n+1}%
\sum _{l=1}^{\infty}\Bigg[\nonumber\\
&&\sum _{h=1}^l(2h+1)
(2l-2h+3)\nonumber\\
&&c_{2h+1}c_{2l-2h+3}\Bigg]\eta ^{2l+3},
\eea
\bea
&&\frac{a}{n+1}\left(5n+4+\frac{n}{k}\right)m\frac{dm}{d\eta}=\frac{a}{n+1}%
\left(5n+4+\frac{n}{k}\right)\nonumber\\
&&\sum _{l=1}^{\infty}\left[\sum
_{h=1}^l(2l-2h+3)c_{2h+1}c_{2l-2h+3}\right]\eta ^{2l+3}.
\eea

By substituting all the above results in Eq. (\ref{kk1}) we obtain
\begin{eqnarray}
&&\sum_{l=1}^{\infty }\Bigg\{2l(2l+3)c_{2l+3}+a\sum_{h=1}^{l}\Bigg\{\left(
2l-2h+3\right) \nonumber\\
&&\Bigg[ h\left( \frac{2(k+1)n}{n+1}+4\right) +\frac{(k+1)^{2}n%
}{k(n+1)}-4l\Bigg] \times  \notag \\
&&c_{2h+1}c_{2l-2h+3}+\frac{n}{n+1}\sum_{i=1}^{l-h+1}(2l-2h-2i+3)\nonumber\\
&&\left[
(2i+1)k\alpha _{2h}^{+}+\frac{1}{k}\alpha _{2h}^{-}\right]
c_{2i+1}c_{2l-2h-2i+3}\Bigg\}\Bigg\}=0.  \notag  \label{ks} \\
\end{eqnarray}%
From Eq.~(\ref{ks}) the recursive relation between the coefficients $%
c_{2l+1} $ immediately follows, and thus we obtain Eq.~({\ref{ff}). This
ends the proof of \textbf{Theorem 4}. }

The values of the first seven coefficients of the exact power series
solution of the general relativistic stars with arbitrary polytropic index
are presented in Appendix \ref{app}.

\subsection{Applications: Polytropic stars with index $n=1/2$, $n=1/5$, and $%
n=3$}

In the following we present some direct applications of \textbf{Theorem 4},
by comparing the estimations obtained from the exact general power series solution of
Eq.~(\ref{kk}), as given by \textbf{Theorem 4}, with the solution obtained by numerically integrating the structure equations of the star.

\subsubsection{The case $n=1/2$}

As a first application of the exact power series solution of the
relativistic mass equation for polytropic stars we present in detail the
case $n=1/2$. Then Eq.~(\ref{kk}) is given by
\begin{eqnarray}
\hspace{-0.7cm}&&3\eta ^{2}\left[ 1-2a\frac{m(\eta )}{\eta }\right] \frac{dm}{d\eta }\frac{%
d^{2}m}{d\eta ^{2}}+\left[ 13am(\eta )-6\eta \right] \left( \frac{dm}{d\eta }%
\right) ^{2}\nonumber\\
\hspace{-0.7cm}&&+a\eta \left( \frac{dm}{d\eta }\right) ^{3}+\frac{ak}{\eta ^{3}}%
\left( \frac{dm}{d\eta }\right) ^{5}+  \frac{a}{k}\eta ^{4}m(\eta )=0.  \label{qa}
\end{eqnarray}
Then the coefficients of the general power series solution of Eq.~({\ref{qa}%
) of the form $m(\eta )=\sum_{l=1}^{\infty }c_{2l+1}\eta ^{2l+1}$ can be
obtained immediately from \textbf{Theorem 4}, and are given by
\begin{equation}
c_{5}=-\frac{a(k+1)(3k+1)}{90k},
\end{equation}%
\begin{equation}
c_{7}=\frac{a^{2}(k+1)(3k+1)\left( 30k^{2}-11k-1\right) }{11340k^{2}},
\end{equation}%
\bea
c_{9}&=&-\frac{a^{3}(k+1)(3k+1)}{2755620k^{3}}\Bigg(
2205k^{4}-1131k^{3}+\nonumber\\
&&620k^{2}+45k+9\Bigg) ,
\eea
\bea
c_{11}&=&\frac{a^{4}(k+1)(3k+1)}{
1818709200k^{4}}\Bigg(
496125k^{6}-304515k^{5}+\nonumber\\
&&263424k^{4}-108691k^{3}-58k^{2}-3582k-279\Bigg) ,\nonumber\\
\eea
\begin{eqnarray}
c_{13} &=&-\frac{a^{5}(k+1)(3k+1)}{2127889764000k^{5}}\Bigg(%
212837625k^{8}-\nonumber\\
&&148281300k^{7}+
160686045k^{6}-\nonumber\\
&&102260088k^{5}+42372329k^{4}-
4403892k^{3}+\nonumber\\
&&2118955k^{2}+305712k+17478\Bigg)%
,
\end{eqnarray}%
\begin{eqnarray}
c_{15} &=&\frac{a^{6}(k+1)(3k+1)}{2010855826980000k^{6}}\Bigg(%
77472895500k^{10}-\nonumber\\
&&59686119675k^{9}+
74882398185k^{8}-\nonumber\\
&&58288209489k^{7}+
36831147573k^{6}-\nonumber\\
&&15322471612k^{5}+3151243758k^{4}-\nonumber\\
&&1002405221k^{3}-186941375k^{2}-\nonumber\\
&&21156579k-962217\Bigg).
\end{eqnarray}%

The comparison of the results obtained by numerically solving the structure
equations of the $n=1/2$ polytropic general relativistic star and of the
results obtained from the power series solution, truncated at $l=7$, are
presented in Fig.~\ref{fig4}.

\begin{figure*}[tbp]
\includegraphics[width=7.5cm, angle=0]{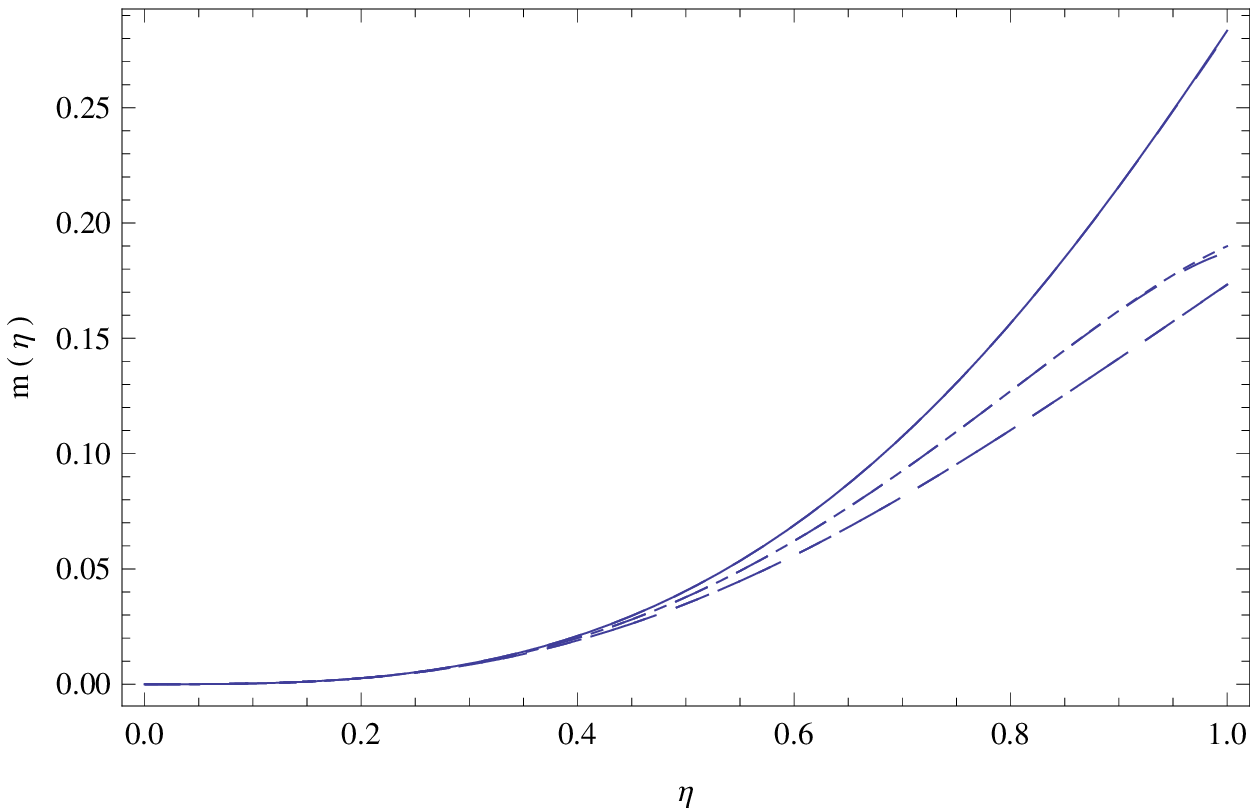} %
\includegraphics[width=7.5cm, angle=0]{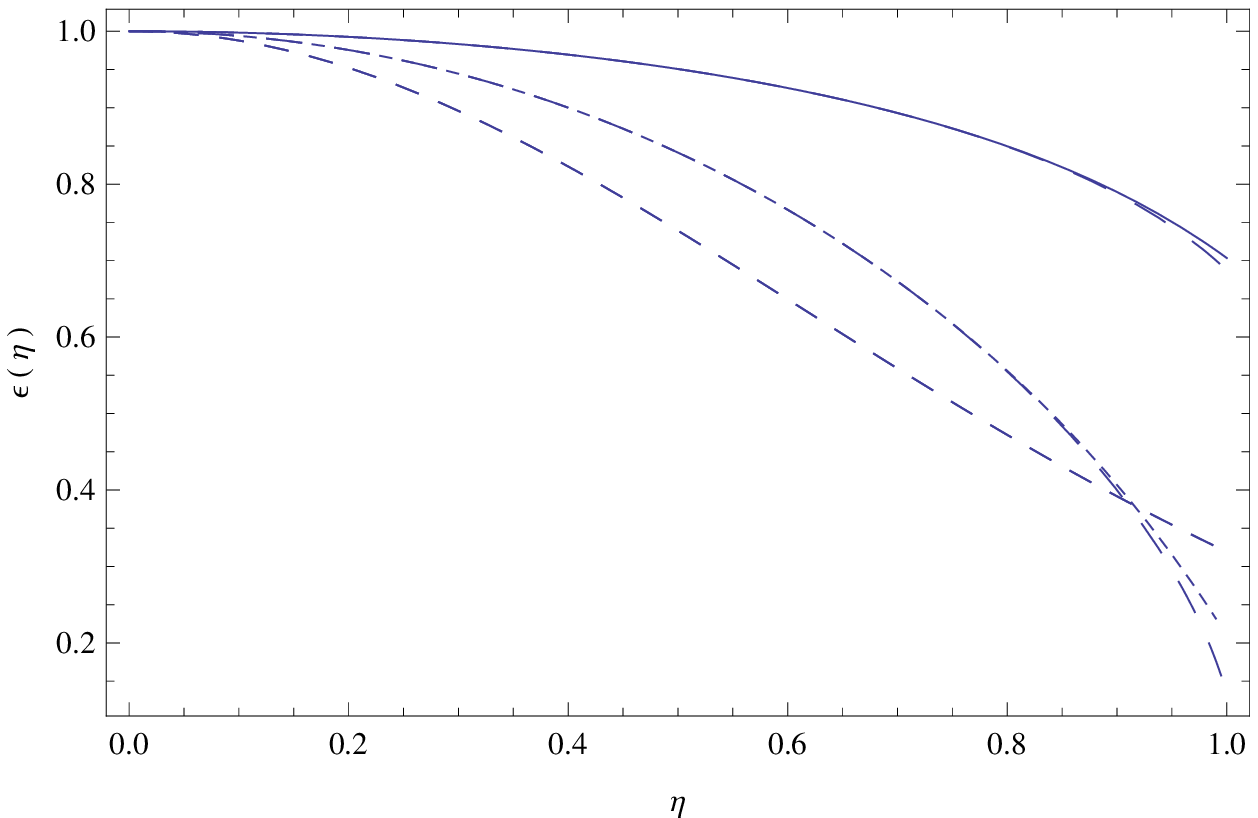}
\caption{Variation with respect to the dimensionless radial coordinate $%
\protect\eta $ of the dimensionless mass $m (\protect\eta)$ (left panel),
and of the dimensionless energy density $\protect\epsilon (\protect\eta )$
(right panel) for a polytropic stars with indexes $n=1/2$, $n=1/5$, and $n=3$, respectively. For the constant $k$ in all cases we have adopted the value  $k=0.1$. In all cases the
long dashed curves represent the numerical solutions of the TOV and mass
continuity Eqs.~(\ref{mcad}) and (\ref{kk}).  The power series solutions of the relativistic mass
equation, truncated to $l=7$, are represented by a solid curve ($n=1/2$), a dotted curve ($n=1/5$), and by a short dashed curve ($n=3$), respectively. }
\label{fig4}
\end{figure*}

For the surface density of the star we have
adopted the value $\epsilon _S=0.209255$, which gives for $a$ the value $%
a=0.76593$. The maximum value of the dimensionless mass $m(1)$ is $%
m(1)=0.189948$. The physical parameters of the star are given by
\begin{equation}
R=9.06873\times \left( \frac{\rho _{c}}{10^{15}\;\mathrm{g/cm^{3}}}\right)
^{-1/2}\;\mathrm{km},
\ee
\be
M_S=0.890139\times \left( \frac{\rho _{c}}{10^{15}\;%
\mathrm{g/cm^{3}}}\right) ^{-1/2}\;M_{\odot }.
\end{equation}

\subsubsection{The case $n=1/5$}

For $n=1/5$, the relativistic mass Eq.~(\ref{p4}) or Eq.~(\ref{kk}) becomes
\begin{eqnarray}\label{x}
\hspace{-0.5cm}&&6\eta ^{2}\left[ 1-2a\frac{m(\eta )}{\eta }\right] \left( \frac{dm}{d\eta }%
\right) ^{4}\frac{d^{2}m}{d\eta ^{2}}+\left[ 25am(\eta )-12\eta \right]\times \nonumber\\
\hspace{-0.5cm}&&\left( \frac{dm}{d\eta }\right) ^{5}+a\eta \left( \frac{dm}{d\eta }\right)
^{6}+  \frac{ak}{\eta ^{9}}\left( \frac{dm}{d\eta }\right) ^{11}+\nonumber\\
\hspace{-0.5cm}&&\frac{a}{k}\eta
^{10}m(\eta )=0.
\end{eqnarray}

We will not present here the explicit expressions of the coefficients $%
c_{2l+1}$ of the exact series solution of Eq.~(\ref{x}). The comparison of
the exact numerical solution of Eq.~(\ref{x}) and of the power series
solution, truncate at $l=7$ is presented in Fig.~\ref{fig4}. For the surface
density of the star we obtain the value $\epsilon _S=0.703388$,
corresponding to $a=0.45734$. The total dimensionless mass of the star is $%
m(1)=0.28339$, giving for the physical parameters of the star the values

\begin{equation}
R=7.00764\times \left( \frac{\rho _{c}}{10^{15}\;\mathrm{g/cm^{3}}}\right)
^{-1/2}\;\mathrm{km},
\ee
\be
M_S=0.612749\times \left( \frac{\rho _{c}}{10^{15}\;%
\mathrm{g/cm^{3}}}\right) ^{-1/2}\;M_{\odot }.
\end{equation}

\subsubsection{The case $n=3$}

A polytrope of the order of $n=3$, corresponding to the equation of state $%
p={\rm constant}\times \rho ^{4/3}$, is known as the Eddington approximation \citep{Edd}.  From
an astrophysical point of view it corresponds to a wholly radiative star.
Thus, for example, the $n=3$ polytrope is used to model the astrophysical
properties of our Sun \citep{Sun}. For $n=3$ the relativistic mass equation
takes the form
\bea
\hspace{-0.5cm}&&\eta ^{2}\frac{d^{2}m}{d\eta ^{2}}-2\frac{dm}{d\eta }+\nonumber\\
\hspace{-0.5cm}&&\frac{3a}{4k}\frac{%
\left[ m+k\eta ^{3}\left( m^{\prime }/\eta ^{2}\right) ^{4/3}\right] \left[
m^{\prime }+k\eta ^{2}\left( m^{\prime }/\eta ^{2}\right) ^{4/3}\right] }{%
\eta \left( 1-2am/\eta \right) \left( m^{\prime }/\eta ^{2}\right) ^{1/3}}=0, \nonumber\\
\label{ne3}
\eea
where we have used the relation $dP/d\epsilon =(4/3)k\epsilon ^{1/3}$, giving $P'\left(m'(\eta)/\eta ^2\right)=(4k/3)\left(m'/\eta ^2\right)^{1/3}$.

The coefficients of the power series solutions of Eq.~(\ref{ne3}) are given
by
\begin{equation}
c_{5}=-\frac{a(k+1)(3k+1)}{40k},
\end{equation}%
\begin{equation}
c_{7}=\frac{a^{2}(k+1)(3k+1)\left( 105k^{2}+34k+19\right) }{13440k^{2}},
\end{equation}%
\bea
c_{9}&=&-\frac{a^{3}(k+1)(3k+1)}{8709120k^{3}}\Bigg(
24255k^{4}+10374k^{3}+\nonumber\\
&&7390k^{2}+470k+619\Bigg) ,
\eea
\begin{eqnarray}
c_{11} &=&\frac{a^{4}(k+1)(3k+1)}{5109350400k^{4}}\Bigg(%
5457375k^{6}+2399670k^{5}+\nonumber\\
&&2059023k^{4}+ 82268k^{3}+244509k^{2}-58914k+ \notag \\
&&
17117\Bigg),
\end{eqnarray}%
\begin{eqnarray}
c_{13} &=&-\frac{a^{5}(k+1)(3k+1)}{7970586624000k^{5}}\Bigg(%
3421774125k^{8}+ \notag \\
&&
1377441450k^{7}+
1447118820k^{6}-29503578k^{5}+ \notag \\
&&
227366374k^{4}-89104202k^{3}+35760180k^{2}- \notag \\
&&
13335878k+1208293\Bigg),
\end{eqnarray}%
\begin{eqnarray}
c_{15} &=&\frac{a^{6}(k+1)(3k+1)}{40171756584960000k^{6}}\Bigg(%
7161773243625k^{10}+ \notag \\
&&
2400911008650k^{9}+  3276987869295k^{8}- \notag \\
&&
339209275608k^{7}+684186777306k^{6}- \notag \\
&&327746282804k^{5}+
\notag \\
&&171863068046k^{4}-75604400072k^{3}+ \notag \\
&&27410294125k^{2}-5942307478k+  \notag \\
&&267910291\Bigg).
\end{eqnarray}

In Fig.~\ref{fig4} we compare the power series solution for the polytropic index $n=3$, truncated to seven
terms, with the exact numerical solution. For $%
a$ we have adopted the value $a=0.687329$, giving a surface density $%
\epsilon _{S}=0.319079$. The total dimensionless mass of the star is $%
m(1)=0.173226$. The physical parameters of the compact general relativistic
object described by the $n=3$ polytrope can be obtained as
\begin{equation}
R=8.59081\times \left( \frac{\rho _{c}}{10^{15}\;\mathrm{g/cm^{3}}}\right)
^{-1/2}\;\mathrm{km},
\ee
\be
M_{S}=0.69008\times \left( \frac{\rho _{c}}{10^{15}\;%
\mathrm{g/cm^{3}}}\right) ^{-1/2}\;M_{\odot }.
\end{equation}

\section{Conclusions and final remarks}\label{sect5}

In the present paper we have obtained exact power series solutions of the
mass continuity and hydrostatic equilibrium equation describing the
structure of general relativistic stars. In order to obtain the solutions we
have formulated the second order differential equation describing the mass
profile of the stars. The relativistic mass equation admits exact,
convergent and non-singular, power series solutions for both the linear
barotropic and polytropic equations of state. We have obtained the power
series solutions for arbitrary values of $\gamma $ for the linear barotropic
equation of state $p=\gamma \rho c^{2}$, and for the general case of the
arbitrary polytropic index $n$. We have compared in detail our exact results
with the results obtained by numerically integrating the gravitational field
equations, by considering the cases $\gamma =1/3$, $\gamma =1$, and $%
n=1,1/2,1/5,3$, respectively. By truncating our power series to only seven
terms we can basically reproduce the numerical results for the mass and
density distribution of the general relativistic stars described by linear
barotropic and polytropic equations of state. The power series solution are
non-singular at the center of the star, and they can be extended up to the
vacuum boundary/surface of the dense matter distribution. Due to the adopted
equations of state the physical requirements for the acceptability of the
solutions are automatically satisfied. Thus, the speed of sound $c_{s}=\sqrt{%
\partial p/\partial \rho }=\sqrt{\gamma }c$ is a constant inside the star,
and for $\gamma \in \lbrack 0,1]$ satisfies the constraint $c_{s}\leq c$.

For the polytropic stars we obtain
\be
c_{s}=\sqrt{k(n+1)/n}\rho ^{1/2n}=c_{s}\left( \rho _{c}\right) \epsilon
^{1/2n},
\ee
where we have denoted
\be
c_{s}\left( \rho _{c}\right) =\rho
_{c}^{1/2n}\sqrt{k(n+1)/n}.
\ee
Using the relation $\epsilon \left(
\eta \right) =\sum_{i=1}^{\infty }\left( 2i+1\right) c_{2i+1}\eta ^{2i-2}$ then we find
\be
c_{s}=c_{s}\left( \rho _{c}\right) \Bigg[
\sum_{i=1}^{\infty }{\left( 2i+1\right) c_{2i+1}\eta ^{2i-2}}\Bigg] ^{
1/2n}\leq c.
\ee

The power series solutions of the Newtonian Lane-Emden Eq.~(\ref{F1}) have
been intensively studied in the astrophysical and mathematical literature
\citep{CA,IW,CH,MN}. The series solutions are represented as $\theta =\sum
_{k=0}^{\infty}{a_k\xi ^{2k}}$ and $\theta ^n= \sum _{k=0}^{\infty}{b_k\xi
^{2k}}$, respectively, with $a_0=b_0=1$ \citep{MN}. One can define the radius
of convergence of these series as the distance from $\xi = 0$ to the closest
singularity of $\theta (\xi)$ in the complex $\xi $-plane. Non-linear
ordinary differential equations, such as the Lane–Emden
equation for $n>1$, can have two kinds of singularities, fixed and movable
\citep{MN}. The Lane–Emden equation for polytropic index $n>1$
and its $n\rightarrow \infty$ limit, corresponding to the limit of the
isothermal sphere equation, are singular at some negative value of the
radius squared \citep{MN}. It is this singularity that prevents real power
series solutions about the center to converge to the outer surface once the
condition $n>1.9121$ is satisfied. However, as shown in \cite{MN}, an Euler
transformation gives power series that do converge up to the outer radius.
Moreover, the Euler-transformed series converge significantly faster than
the series obtained in \cite{IW}, which are limited to finite radii whenever
$n>5$ by a complex conjugate pair of singularities. Series solutions for
polytropic stars by using the Euler transform were constructed in \cite{MN},
so that longer than 60-term series are needed for the outer regions of $n>3$
polytropic Newtonian stars, while 120-term and 300-term series are needed to
obtain the function $\theta (\xi)$ to seven decimal place accuracy all the
way from the center to the surface of the compact object for $n=3.5$ and $%
n=4 $, respectively \citep{MN}. In this context we would like to point out
that the power series solutions of the relativistic mass equation can be
extended to the boundary of the considered stars, and only seven terms are
required to reproduce the numerical solutions with a high precision.

Although the numerical solutions of the structure equations of spherically
symmetric static general relativistic stars can be obtained numerically in a
very efficient, simple and accurate way, we must point out that power series
represent one of the most powerful methods of mathematical analysis. The use
of power series is no less convenient than the use of elementary functions,
especially when solutions of differential equations are to be studied
numerically. In the case of the approach based on the relativistic mass
equation an important advantage of a power series solution is that it gives
the value of the mass and energy density inside the star as a recurrent
power series in the radial coordinate $r$, since the dimensionless variable $%
\eta =r/R$. Consequently, we can predict the physical and geometrical
parameters of the star at any radius directly. Moreover, power series
analytical solutions describing the interior of compact general relativistic
objects usually offers deeper insights into their physical and geometrical
properties, thus offering the possibility of a better understanding of the
structure of dense stars.

\section*{Acknowledgments}

We would like to thank to the anonymous referee for comments and suggestions that helped us to significantly improve our manuscript.

\appendix

\section{The first seven coefficients of the power series solution of the
relativistic mass equation for arbitrary polytropic index $n$}\label{app}

The first seven coefficients $c_{2l+1}$, $l=1,2,...,7$ describing the
solution of the relativistic mass equation for a general relativistic
polytropic star with arbitrary polytropic index $n\in \mathbf{R}$ are given
by
\begin{equation}
c_3=\frac{1}{3},
\end{equation}
\begin{equation}
c_{5}=-\frac{a(k+1)(3k+1)n}{30k(n+1)},
\end{equation}%
\begin{equation}
c_{7}=\frac{a^{2}(k+1)(3k+1)n}{2520k^{2}(1+n)^{2}}\left[
k^{2}(30n+15)+k(18n-20)+8n-5\right] ,
\end{equation}%
\begin{eqnarray}
c_{9} &=&-\frac{a^{3}(1+k)(1+3k)n}{408240k^{3}(1+n)^{3}}\Bigg[315k^{4}\left(
6n^{2}+7n+2\right) +6k^{3}\left( 288n^{2}-241n-140\right) +  \notag \\
&&2k^{2}\left( 618n^{2}-809n+560\right) +10k\left( 40n^{2}-123n+56\right)
+122n^{2}-183n+70\Bigg],
\end{eqnarray}%
\begin{eqnarray}
c_{11} &=&\frac{a^{4}(k+1)(3k+1)n}{179625600k^{4}(n+1)^{4}}\Bigg[%
14175k^{6}\left( 24n^{3}+46n^{2}+29n+6\right) +90k^{5}\Big(%
4074n^{3}-1727n^{2}-  \notag \\
&&4402n-1260\Big)+k^{4}\left( 323568n^{3}-412518n^{2}+330915n+160650\right)
+4k^{3}  \notag \\
&&\left( 38832n^{3}-139547n^{2}+106520n-50400\right) +k^{2}\left(
71744n^{3}-256154n^{2}+418725n-154350\right) +  \notag \\
&&18k\left( 942n^{3}-5847n^{2}+6490n-2100\right)
+5032n^{3}-12642n^{2}+10805n-3150\Bigg],
\end{eqnarray}%
\begin{eqnarray}
c_{13} &=&-\frac{a^{5}(k+1)(3k+1)n}{70053984000k^{5}(n+1)^{5}}\Bigg[%
467775k^{8}\left( 120n^{4}+326n^{3}+329n^{2}+146n+24\right) +1350k^{7}\times
\notag \\
&&\left( 47748n^{4}+2516n^{3}-77423n^{2}-55548n-11088\right) +180k^{6}\Big(%
354618n^{4}-343910n^{3}+  \notag \\
&&333441n^{2}+518150n+124740\Big)+6k^{5}\Big(%
5942244n^{4}-22880944n^{3}+17820615n^{2}-  \notag \\
&&11274200n-4851000\Big)+2k^{4}\Big(%
9945804n^{4}-43963854n^{3}+95782105n^{2}-52438750n+  \notag \\
&&17740800\Big)+2k^{3}\left(
3358788n^{4}-27242948n^{3}+54777285n^{2}-60621700n+18711000\right) +  \notag
\\
&&20k^{2}\left( 138430n^{4}-873842n^{3}+2695909n^{2}-2399130n+679140\right)
+k\Big(495624n^{4}-  \notag \\
&&5988304n^{3}+11652870n^{2}-8520800n+2217600\Big)+183616n^{4}-663166n^{3}+
\notag \\
&&915935n^{2}-574850n+138600\Bigg],
\end{eqnarray}%
\begin{eqnarray}
c_{15} &=&\frac{a^{6}(k+1)(3k+1)n}{88268019840000k^{6}(n+1)^{6}}\Bigg[%
42567525k^{10}\Big(720n^{5}+2556n^{4}+3604n^{3}+2521n^{2}+874n+  \notag \\
&&120\Big)+28350k^{9}\left(
1249692n^{5}+686860n^{4}-2577633n^{3}-3421416n^{2}-1539028n-240240\right) +
\notag \\
&&135k^{8}\Big(%
278662344n^{5}-141122820n^{4}+282670258n^{3}+825130367n^{2}+466600470n+
\notag \\
&&79879800\Big)+36k^{7}\Big(%
612790998n^{5}-2472456073n^{4}+1527229265n^{3}-1345472610n^{2}-  \notag \\
&&1877749300n-399399000\Big)+6k^{6}\Big(%
2273834244n^{5}-10928033994n^{4}+29030321910n^{3}-  \notag \\
&&15039510755n^{2}+7751107700n+3006003000\Big)+4k^{5}\Big(%
1272432138n^{5}-12749058409n^{4}+  \notag \\
&&32186860166n^{3}-50094341110n^{2}+21616077000n-5381376000\Big)+2k^{4}\Bigg(%
1245531860n^{5}-  \notag \\
&&10585913054n^{4}+47107629124n^{3}-68141956465n^{2}+56246234100n-14777763000%
\Bigg)+  \notag \\
&&4k^{3}\Big(%
144913746n^{5}-2616936275n^{4}+9372735679n^{3}-19092417410n^{2}+14124980100n-
\notag \\
&&3552549000\Big)+5k^{2}\Big(%
58083800n^{5}-526446912n^{4}+2878464312n^{3}-4529906063n^{2}+  \notag \\
&&2903305230n-685284600\Big)+6k\Big(%
5002696n^{5}-154497310n^{4}+462711169n^{3}-  \notag \\
&&560651580n^{2}+316735300n-70070000\Big)%
+21625216n^{5}-103178392n^{4}+200573786n^{3}-  \notag \\
&&199037015n^{2}+101038350n-21021000\Bigg].  \label{c15}
\end{eqnarray}


\begin{thebibliography}{}

\bibitem[Arba\~{n}il et al.(2013)]{jj} Arba\~{n}il, J. D. V., Lemos,  J. P. S.,  \& Zanchin, V. T., 2013,
Phys. Rev. D, 88, 084023

\bibitem[Binney \& Tremaine(1987)]{2s} Binney, J. \&Tremaine, S., 1987, Galactic dynamics, Princeton, Princeton
University Press, U. S. A.

\bibitem[Bahcall \& Ulrich(1988)]{Sun} Bahcall,  J. N. \& Ulrich, R. K., 1988,  Rev. Mod. Phys., 6,
297

\bibitem[Bhar(2015)]{B2} Bhar, P., 2015, Astrophys. Space Sci., 356, 309

\bibitem[Boehmer \& Harko(2007)]{BoHa07} Boehmer,  C. G. \& Harko, T., 2007, Journal of Cosmology and Astroparticle
Physics, 0706, 025

\bibitem[Boehmer \& Harko(2010)]{KCC} Boehmer,  C. G. \& Harko, T., 2010, Journal of Nonlinear Mathematical Physics,
17, 503

\bibitem[Buchdahl(1959)]{Bu} Buchdahl, H. A., 1959,  Phys. Rev., 116, 1027

\bibitem[Burikham et al.(2015)]{Pi1} Burikham, P., Cheamsawat, K., Harko, T., \& Lake, J. M., 2015,  Eur. Phys. J. C, 75, 442  

\bibitem[Burikham et al.(2016a)]{Pi2} Burikham, P., Cheamsawat, K., Harko, T., \& Lake, J. M., 2016,  Eur. Phys. J. C, 76, 106

\bibitem[Burikham et al.(2016b)]{Pi3} Burikham, P., Harko, T., \& Lake, J. M., 2016,  eprint arXiv:1606.05515

\bibitem[Chandrasekhar(2012)]{1s} Chandrasekhar, S., 2012, An introduction to the study of stellar
structure, Dover Books on Astronomy Series, Dover Publications, Dover, United Kingdom

\bibitem[Chang \& Mott(1975)]{CM} Chang, F. C. \& Mott, H., 1975, Journal of The Franklin Institute, 299,
227

\bibitem[Chavanis(2008)]{PH} Chavanis, P. H., 2008, Astron. \& Astrophys., 483, 673

\bibitem[Chavanis \& Harko(2012)]{ChHa} Chavanis,  P. H. \& Harko, T., 2012, Phys. Rev. D, 86, 064011

\bibitem[Delgaty \& Lake(1998)]{1b}  Delgaty, M. S. R. \& Lake, K., 1998, Comput. Phys. Commun., 115, 395
(1998).

\bibitem[Dev \& Gleiser(2003)]{B4} K. Dev, K. \& Gleiser, M., 2003, Gen. Rel. Grav., 35, 1435

\bibitem[Dev \& Gleiser(2004)]{B3} Dev, K. \&  Gleiser, M., 2004, Int. J. Modern Phys. D, 13, 1389

\bibitem[Eddington(1926)]{Edd} Eddington, A. S., 1926, The Internal Constitution of the Stars, Cambridge, Cambridge University Press

\bibitem[Emden(1907)]{E} Emden, R., 1907, Gss balls: Applications of the Mechanical Heat Theory to Cosmological and Meteorological Problems (in German), Teubner, Berlin

\bibitem[Finch \& Skea(1998)]{2b}  Finch, M. R., \&  Skea, J. E. F., 1998, unpublished preprint,
www.dft.if.uerj.br/users/Jim Skea/papers/pfrev.ps

\bibitem[Fiziev \& Marinov(2015)]{PK}  Fiziev, P. \& Marinov, K., 2015, Bulgarian
Astronomical Journal, 23, 3

\bibitem[Folomeev \& Singleton(2012)]{VD} Folomeev, V. \& Singleton, D., 2012, Phys. Rev. D, 85, 064045

\bibitem[Fowler(1930)]{Fow} Fowler, R. H., 1930, Mon. Not. R. Astron. Soc., 91, 63

\bibitem[Herrera \& Barreto(2013)]{LW}  Herrera, L. \&  Barreto, W., 2013, Phys.
Rev. D, 88, 084022

\bibitem[Horedt(2004)]{Hor} Horedt,  G. P., 2004, Polytropes. Applications in Astrophysics and
Related Fields, Kluwer Academic Publishers, Dordrecht, The Netherlands

\bibitem[Hunter(2001)]{CH} Hunter, C., 2001, Mon. Not. R. Astron. Soc., 328, 839

\bibitem[Kippenhahn \& Weigert(1990)]{3s} Kippenhahn,  R. \& Weigert, A., 1990,  Stellar structure and evolution,
Berlin, Springer-Verlag, Germany

\bibitem[Kramer et al.(1980)]{0} Kramer, D., Stephani, H., MacCallum, M., \&  Herlt, E., 1980, Exact
solutions of Einstein's field equations, Cambridge, Cambridge University
Press

\bibitem[Lai \& Xu(2009)]{XY} Lai, X. Y. \& Xu, R. X., 2009,
Astropart. Phys., 31, 128

\bibitem[Landau \& Lifshitz(1975)]{Landau} Landau, L. D. \& Lifshitz, E. M., 1975,  The classical theory of fields,
Butterworth-Heinemann, Oxford, United Kingdom

\bibitem[Lane(1870)]{L} Lane, J. H., 1870,  The American Journal of Science and Arts, 50, 57

\bibitem[Mach(2012)]{PM} Mach, P., 2012, Journal of Mathematical Physics, 53, 062503

\bibitem[Mafa Takisa \& Maharaj(2013)]{PS} Mafa Takisa, P. \& Maharaj, S. D., 2013, Gen. Rel. Grav., 45, 1951

\bibitem[Mak et al.(2000)]{MCC}  Mak, M. K., Dobson Jr, P. N., \& Harko, T., 2000, Mod. Phys.Lett. A, 15, 2153 

\bibitem[Mak et al.(2002a)]{A1} Mak, M. K.,  Dobson Jr, P. N., \& Harko, T., 2002, Int. J. Modern Phys. D, 11, 207

\bibitem[Mak \& Harko(2002b)]{A3}  Mak, M. K. \& Harko, T., 2002,  Annalen Phys. (Berlin), 11, 3

\bibitem[Mak \& Harko(2002c)]{A4}  Mak, M. K. \& Harko, T., 2002, Chin. J. Astronomy Astrophysics, 2, 248

\bibitem[Mak \& Harko(2003)]{A2} Mak, M. K. \& Harko, T., 2003, Proc. Royal Soc. Lond. A, 459, 393

\bibitem[Mak \& Harko(2004)]{BV} Mak, M. K. \& Harko, T., 2004, Int. J. Mod. Phys. D, 13,
149

\bibitem[Mak \& Harko(2005)]{MMH} Mak, M. K. \& Harko, T., 2005, Pramana, 65, 185

\bibitem[Mak \& Harko(2012)]{M1} Mak, M. K. \& Harko, T., 2012, Applied Mathematics and Computations, 218,
10974

\bibitem[Mak \& Harko(2013a)]{M2} Mak, M. K. \& Harko, T., 2013, Applied Mathematics and Computations,
219, 7465

\bibitem[Mak \& Harko(2013b)]{00} Mak, M. K. \& Harko, T., 2013, European Physical J C, 73, 2585

\bibitem[Mancas \& Rosu(2016)]{Mancas} Mancas, S. C. \& Rosu, H. C., 2016,  arXiv:1604.04807

\bibitem[Maurya et al.(2015)]{B1}  Maurya, S. K.,  Gupta, Y. K.,  Ray, S.,  \& Dayanandan, B., 2015,
European Physical J C, 75, 225

\bibitem[Milne(1930)]{Milne} Milne, E. A., 1930, Mon. Not. R. Astron. Soc., 91, 4

\bibitem[Mitra \& Glendenning(2010)]{Glen}  Mitra, A. \& Glendenning, N. K., 2010, Monthly Notices of the Royal Astronomical Society: Letters,
404, L50

\bibitem[Mohan \& Al-Bayati(1980)]{CA} Mohan, C. \& Al-Bayaty, A. R., 1980,  Astrophys. Space Sci., 73, 227

\bibitem[Nilsson \& Uggla(2000)]{UN} Nilsson, U. S. \&  Uggla, C., 2000, Annals of Physics, 286,
292

\bibitem[Nouh(2004)]{MN} Nouh,  M. I., 2004,  New Astronomy, 9, 467

\bibitem[Nouh \& Saad(2013)]{MAS} Nouh,  M. I. \& Saad, A. S., 2013,  International Review of
Physics, 7, 16

\bibitem[Olmo(2008)]{GO} Olmo, G. J., 2008,  Phys. Rev. D, 78, 104026

\bibitem[Picanco et al.(2004)]{RM} Picanco, R., Malheiro, M.,  \&  Ray, S., 2004,  Int. J. Modern Phys.
D, 13, 1441

\bibitem[Rhoades \& Ruffini(1974)]{Ruf} Rhoades,  C. E. \& Ruffini, R., 1974, Physical Review Letters, 32, 324

\bibitem[Riazi et al.(2015)]{3}  Riazi, N., Sedigheh Hashemi, S,  Naseh Sajadi, S.,  \&  Assyyaee, S., 2015,
arXiv:1507.03420.

\bibitem[Roxburgh \& Stockman(1999)]{IW} Roxburgh, I. W. \& Stockman, L. M., 1999,  Mon. Not. R. Astron. Soc., 303,  466

\bibitem[S\'{a}(1999)]{PTA} S\'{a}, P. M., 1999, Phys. Lett. B, 467, 40

\bibitem[Schmidt \& Homann(2000)]{Sch} Schmidt, H.-J. \&  Homann, F., 2000, Gen. Rel.
Grav., 32, 919

\bibitem[Schwarzschild(1916)]{Sch1} Schwarzschild, K., 1916, Sitzber Deut. Akad.
Wiss. Math. Phys. Berlin, 23, 189

\bibitem[Tolman(1939)]{Tolman} Tolman, R. C., 1939, Phys. Rev., 55, 364

\bibitem[Tooper(1964)]{Tooper} Tooper,  R. F., 1964,  Astrophysical Journal, 140, 434

\bibitem[Wolfram(2003)]{Wolfram} Wolfram, S., 2003, The Mathematica Book, 5th edition, Wolfram Media, Champaigne, USA

\end{thebibliography}
\end{document}